\documentclass[onecolumn, draftclsnofoot,11pt]{IEEEtran}
\IEEEoverridecommandlockouts

\usepackage{mathrsfs}
\usepackage{tikz}
\usepackage[utf8]{inputenc}
\usepackage{amsfonts} 
\usepackage{epstopdf}
\usepackage{dsfont}
\usepackage{graphicx}
\usepackage{bbm}
\usepackage{enumerate}
\usepackage{epsf,epsfig,verbatim,amssymb,amsmath,array,cite,amsthm,subfigure,multicol,multirow}  
\usepackage{psfrag,xspace}

\usepackage{bm}
\usepackage{amssymb}
\usepackage{hhline}
\usepackage{mathabx}
\usepackage{physics}
\usepackage{xcolor}
\definecolor{darkblue}{rgb}{0.1,0.1,0.8}
\definecolor{brickred}{rgb}{0.8, 0.25, 0.33}
\definecolor{DarkGreen}{rgb}{0,0.6,0}
\newtheorem{theorem}{Theorem}

\theoremstyle{definition}
\newtheorem{definition}{Definition}
\newtheorem{example}{Example}

\newtheorem{prop}{Proposition}
\newtheorem*{prob*}{Problem}

\newtheorem{remark}{Remark}

\allowdisplaybreaks 
\makeatletter
\def\old@comma{,}
\catcode`\,=13
\def,{%
  \ifmmode%
    \old@comma\discretionary{}{}{}%
  \else%
    \old@comma%
  \fi%
}
\makeatother

\usepackage{etoolbox}
\providetoggle{Blue_revision}
\settoggle{Blue_revision}{true}

 \usepackage{hyperref}
\hypersetup{
colorlinks=true, %
 pdfstartview={FitH},
    linkcolor=red,
    citecolor=blue, 
    urlcolor={blue!80!black}
}

\usepackage{graphicx}
\usepackage{stmaryrd }
\usepackage{stackengine}

\usepackage{relsize}
\usepackage{upgreek}

\global\long\def\CC{\mathbb{C}}

\global\long\def\EE{\mathbb{E}}

\global\long\def\FF{\mathbb{F}}
\global\long\def\11{\mathbbm{1}}

\def\CalA{\mathcal{A}}

\def\CalD{\mathcal{D}}

\def\CalH{\mathcal{H}}

\def\CalN{\mathcal{N}}

\def\CalP{\mathcal{P}}
\def\CalQ{\mathcal{Q}}
\def\CalR{\mathcal{R}}
\def\CalS{\mathcal{S}}
\def\CalW{\mathcal{W}}
\def\CalU{\mathcal{U}}
\def\CalV{\mathcal{V}}
\def\CalT{\mathcal{T}}
\def\CalX{\mathcal{X}}

\global\long\def\+{\oplus}

\usepackage{stackengine}
\def\deq{\mathrel{\ensurestackMath{\stackon[1pt]{=}{\scriptstyle\Delta}}}}

\usepackage{acronym}
\usepackage{enumitem,kantlipsum}

\newacro{qc}[q-c]{quantum-to-classical }

\global\long\def\tensor{\otimes}
 \def\tr{\Tr}

\allowdisplaybreaks

\def\IndiU1{\mathbbm{1}_{\left\{U^{n,(\mu_1)}(l)=u^n\right\}} }
\def\IndiV1{\mathbbm{1}_{\left\{V^{n,(\mu_2)}(k)=v^n\right\}} }

\def\define{\mathrel{\ensurestackMath{\stackon[1pt]{=}{\scriptstyle\Delta}}}}

\def\ulineU{\underline{U}}
\def\ulineV{\underline{V}}
\def\ulineM{\underline{M}}
\def\ulineW{\underline{W}}
\def\ulines{\underline{s}}
\def\ulinea{\underline{a}}
\def\ulineu{\underline{u}}
\def\ulinev{\underline{v}}
\def\ulinem{\underline{m}}
\def\TDelta{\mathcal{T}_{\delta}^{(n)}}
\def\fieldq{\mathbb{F}_{q}}

\def\ulinel{\underline{w}}
\def\ulinew{\underline{w}}

\def\CalN{\mathcal{N}}

\def\HZ{{\mathcal{H}^{\otimes n}_Z}}
\def\HZprime{{\mathcal{H}_{Z}'}^{\otimes n}}
\def\ovec{|0\rangle\langle0|^{\mathbb{C}^{2n}}}
\def\holevoPOVM{\lambda_{(a,m),m_3,m_4}}
\def\avgPeNprime{\bar{\xi}(c^{(n)},\CalN'_4)}
\def\avgPeN{\bar{\xi}(c^{(n)},\CalN_4)}

\def\HZhat{\bar{\mathcal{H}}_Z}

\newcommand{\bbm}[1]{\mathbbm{#1}}
\newcommand{\mcl}[1]{\mathcal{#1}}
\newcommand{\ul}[1]{\underline{#1}}
\newcommand{\h}[1]{\hat{#1}}

\newcommand{\ral}{\rangle}
\newcommand{\lal}{\langle}

\def\Piv{\Pi_{v^n\ulineu^n}^{\mathrm{V}}} 
\def\Piuj{\Pi_{v^n\ulineu^n}^{\mathrm{U_j}}}
\def\Pivuj{\Pi_{v^n\ulineu^n}^{\mathrm{VU_j}}}
\def\Piu{\Pi_{v^n\ulineu^n}^{\mathrm{\underline{U}}}}
\def\Pivu{\Pi_{v^n\ulineu^n}^{\mathrm{V\underline{U}}}}

\def\projv{\bar{\Pi}_{v^n\ulineu^n}^{\mathrm{V}}} 
\def\projuj{\bar{\Pi}_{v^n\ulineu^n}^{\mathrm{U_j}}}
\def\projvuj{\bar{\Pi}_{v^n\ulineu^n}^{\mathrm{VU_j}}}
\def\proju{\bar{\Pi}_{v^n\ulineu^n}^{\mathrm{\underline{U}}}}
\def\projvu{\bar{\Pi}_{v^n\ulineu^n}^{\mathrm{V\underline{U}}}}

\def\spv{{\bar{\Omega}}_{v^n\ulineu^n}^{\mathrm{V}}} 
\def\spuj{{\bar{\Omega}}_{v^n\ulineu^n}^{\mathrm{U_j}}}
\def\spvuj{{\bar{\Omega}}_{v^n\ulineu^n}^{\mathrm{VU_j}}}
\def\spu{{\bar{\Omega}}_{v^n\ulineu^n}^{\mathrm{\underline{U}}}}
\def\spvu{{\bar{\Omega}}_{v^n\ulineu^n}^{\mathrm{V\underline{U}}}}

\def\tspv{{\Omega}_{v^n\ulineu^n w_V^n \ulinew_U^n}^{\mathrm{V}}} 
\def\tspuj{{\Omega}_{v^n\ulineu^n w_V^n \ulinew_U^n}^{\mathrm{U_j}}}
\def\tspvuj{{\Omega}_{v^n\ulineu^n w_V^n \ulinew_U^n}^{\mathrm{VU_j}}}
\def\tspu{{\Omega}_{v^n\ulineu^n w_V^n \ulinew_U^n}^{\mathrm{\underline{U}}}}
\def\tspvu{{\widehat{\Omega}}_{v^n\ulineu^n w_V^n \ulinew_U^n}}

\begin{document}

\title{\huge Unified approach for computing sum of sources over CQ-MAC}

\author{\IEEEauthorblockN{Mohammad Aamir Sohail\IEEEauthorrefmark{1}, Touheed Anwar Atif\IEEEauthorrefmark{1},
		S. Sandeep Pradhan\IEEEauthorrefmark{1}, and Arun Padakandla\IEEEauthorrefmark{2} \\}
	\IEEEauthorblockA{Department of Electrical Engineering and Computer Science,\\
		\IEEEauthorrefmark{1}University of Michigan, Ann Arbor, USA.\\
		\IEEEauthorrefmark{2}University of Tennessee, Knoxville, USA\\
		Email: \tt mdaamir@umich.edu, touheed@umich.edu,  pradhanv@umich.edu,
		arunpr@utk.edu,}
}

\maketitle

\begin{abstract}
  We consider the task of communicating a generic bivariate function of two classical sources over a Classical-Quantum Multiple Access Channel (CQ-MAC). The two sources are observed at the encoders of the CQ-MAC, and the decoder aims at reconstructing a bivariate function from the received quantum state. Inspired by the techniques developed for the analogous classical setting, and employing the technique of simultaneous (joint) decoding developed for the classical-quantum setting, we propose and analyze a coding scheme based on a fusion of algebraic structured and unstructured codes. This coding scheme allows exploiting both the symmetric structure common amongst the sources and the asymmetries. We derive a new set of sufficient conditions that strictly enlarges the largest known set of sources (capable of communicating the bivariate function) for any given CQ-MAC. 
  We provide these conditions in terms of single-letter quantum information-theoretic quantities.
\end{abstract}

{\hypersetup{
colorlinks=true, %
 pdfstartview={FitH},
    linkcolor=black,
    citecolor=blue, 
    urlcolor={blue!80!black}
}
}

\section{Introduction}
\label{sec:introduction}
 In this work, we revisit the problem of computing functions of information sources transmitted over a classical-quantum multiple access channel (CQ-MAC). The problem can be described as follows. Let $( \rho_{x_{1},x_{2}} \in \mathcal{D}(\mathcal{H}_{Y} ) :  (x_{1},x_{2}) \in \mathcal{X}_{1}\times \mathcal{X}_{2})$ be a model for a given CQ-MAC.
Consider a scenario where two distributed parties observe two classical information streams $S_{jt} \in \mathcal{S}_j\colon t\geq1$, with the pair $(S_{1t},S_{2t})\colon t\geq 1$ being independent and identically distributed (IID) according to the distribution $\mathbb{W}_{S_1S_2}$. These parties intend to send a bivariate function of $S_1$ and $S_2$ to a centralized receiver using the above CQ-MAC. The receiver upon receiving the prepared quantum state aims to reconstruct the bivariate function $f$ from the quantum state. In this work, we aim to characterize the sufficient conditions, 
on the distribution of sources $\mathbb{W}_{S_1S_2}$ that for a given CQ-MAC, the centralized decoder reconstructs the bivariate function, with an arbitrary low probability of error. 

The conventional approach to this involves making the receiver to 
completely reconstruct the pair of classical sources, and then characterize the sufficient conditions. This would be a direct consequence of the result derived in \cite{winter2001capacity}. The authors in \cite{hayashi2021computation} address this CQ-MAC problem where the sources are computed directly without the need for the explicit reconstruction of the individual sources. However, they restrict their attention to uniform input distributions.
The authors in \cite{touheed_compCQMAC} instead employed a different technique, using asymptotically good random nested coset codes that directly reconstruct arbitrary function $f$
of sources of arbitrary distributions. Their work was built on the earlier ideas of \cite{korner1979encode,nazer2007computation,padakandla2013computing,ahlswede1983source}, developed for the classical setting, where the authors developed coding techniques allowing the receiver to directly recover the sum of the sources without recovering either of the sources. These techniques are part of a broader framework for the multi-terminal problems, characterized by codes, with asymptotically large block-length and endowed with algebraic structure, achieving performance limits that the conventional techniques based on unstructured random codes cannot \cite{pradhanalgebraic}. 

However, even in the classical multi-terminal setting, the coding techniques relying on the algebraic structure may show gains for only a certain class of problems and in certain rate regimes. 
Therefore, a unified technique that captures the gain of both the traditional unstructured coding techniques and the algebraic structured based techniques is needed to approach the performance limits for the multi-terminal problems. Alhswede-Han \cite{ahlswede1983source} obtained the best known inner bound for the problem of classical lossless distributed compression by combining the Slepian-Wolf \cite{slepian1973noiseless} coding scheme with the algebraic structured based scheme of K\'orner-Marton  \cite{korner1979encode}. 

Motivated by this, 
the main contribution of current work is in providing a unified approach for the problem of computing a bivariate function of two sources over CQ-MAC, capitalizing on the gains of the algebraic structured techniques developed in \cite{touheed_compCQMAC}, while making the most of the standard approach based on unstructured codes developed for this problem \cite{winter2001capacity}.
We propose an approach where each transmitter intends to send two pieces of information about its  corresponding source to the receiver. The first piece of information from both the sources need to be reconstructed individually at the receiver. Then, conditioned on this reconstruction, we let the receiver reconstruct the necessary function $f$ of the second piece. At $i$th transmitter, the two pieces are constructed on auxiliary variables  $U_i$ and $V_i$, and then fused to form the channel input $X_i$. 
We construct a $4-$input CQ-MAC to model this transmission.  
This poses a challenge concerning the number of messages being decoded. 
The decoder aims at decoding the triple $(U_1,U_2,V_1\oplus_q V_2)$, where $\oplus_q$ represents addition 
with respect to a prime finite field $\FF_q$. For this, the decoder needs a CQ simultaneous decoding technique. Although the ideas of joint typicality using tilting, smoothing, and augmentation introduced by Sen \cite{sen2018one,sen2021unions} solved the 
problem of 
simultaneous decoding of individual messages
on CQ-MAC, it is based on unstructured coding techniques. We develop a unified coding framework that combines unstructured and structured coding 
techniques while using the jointly typicality approach of Sen that enable the decoder to reconstruct $(U_1,U_2,V_1 \oplus_q V_2)$
simultaneously. 
 



In light of this, the main contribution of the current work is in providing a new set of sufficient conditions (see Theorems \ref{thm:4-to-3} and \ref{thm:decodingArbitraryfunc}),  while strictly subsuming the current known conditions, for the reconstruction of an arbitrary function of sources over a generic CQ-MAC. We provide these conditions in terms of single-letter quantum information quantities.
Furthermore, we have identified examples where the gains provided by this framework are demonstrated. 
This work opens up the opportunity to investigate 
a generic approach encompassing both the conventional and algebraic structured techniques for other multi-terminal problems in the classical-quantum regime \cite{fawzi2012classical,savov2012network,3userInterference}.

\section{Preliminaries and Notation}
\label{sec:prelim}
\noindent \textbf{Notation:} We supplement the notation in \cite{2013Bk_Wil} with the following. 
For positive integer $n$, $[n] \define \left\{1,\cdots,n \right\}$. 
We employ an \underline{underline} notation to aggregate objects of similar type. For example, $\ulines$ denotes $ (s_{1},s_{2})$, $\underline{x}^{n}$ denotes $(x_{1}^{n},x_{2}^{n})$, $\underline{\mathcal{S}}$ denotes the Cartesian product $\mathcal{S}_{1}\times \mathcal{S}_{2}$.
Let $\mathrm{c.c.}(\mathcal{S}), \FF_{q}, \mbox{ and } \oplus$ denote the convex closure of the set $\CalS$, the unique prime finite field of size $q$, and the addition operation of the prime field $\FF_{q}$, i.e. $\oplus_q$, respectively. 
For a Hilbert space $\mathcal{H}$, $\mathcal{P}(\mathcal{H})$ and $\mathcal{D}(\mathcal{H})$ denote the collection of positive and density operators acting on $\mathcal{H}$, respectively.

\noindent \par Consider a (generic) $2$-user \textit{CQ-MAC} $\mathcal{N}_2$, which is  
specified through (i) finite sets $\mathcal{X}_{j} : j \in [2]$, (ii) Hilbert space $\mathcal{H}_{Y}$, and (iii) a collection $( \rho_{x_{1},x_{2}} \in \mathcal{D}(\mathcal{H}_{Y} ) :  (x_{1},x_{2}) \in \mathcal{X}_{1}\times \mathcal{X}_{2})$ of density operators.
This CQ-MAC is employed to transmit a pair of sources such that the centralized receiver is capable of reconstructing a bivariate function of the classical information streams observed by the senders. Let $\mathcal{S}_{1},\mathcal{S}_{2}$ be finite sets, and let $(S_{1},S_{2}) \in \mathcal{S}_{1}\times \mathcal{S}_{2}$, distributed with PMF $\mathbb{W}_{S_{1}S_{2}}$, model the pair of information sources observed at the encoders. Specifically, sender $j$ observes the sequence $S_{jt} \in \mathcal{S}_{j}: t \geq 1$.  The sequence $(S_{1t},S_{2t}): t \geq 1$ is assumed to be IID with single-letter PMF $\mathbb{W}_{S_{1}S_{2}}$. The receiver aims to recover the sequence $f(S_{1t},S_{2t}) : t \geq 1$ losslessly, where $f:\mathcal{S}_{1}\times \mathcal{S}_{2} \rightarrow \mathcal{R}$ is a specified bivariate function, and $\mathcal{R}$ is some finite set.
\begin{definition}\label{def:CQ-MACcode}
 A CQ-MAC code $c_{f}=(n,e_{1},e_{2},\boldsymbol{\lambda})$ of block-length $n$ for recovering $f$ consists of two encoding maps $e_{j} : \mathcal{S}^{n} \rightarrow \mathcal{X}_{j}^{n} : j \in [2]$, and a POVM $\boldsymbol{\lambda}= \{ \lambda_{r^{n}} \in \mathcal{P}(\mathcal{H}_{Y}) : r^{n} \in \CalR^{n}\}$. The average error probability of the $c_{f}$ is
\begin{align*}
\label{Eqn:ProbStatementGeneralfrrorOfCode}
\overline{\xi}(c_{f}) &= 1- \sum_{\ulines^{n}:f(\ulines^{n})=r^{n}}\mathbb{W}_{S_{1}S_{2}}^{n}(s_{1}^{n},s_{2}^{n})\tr(\lambda_{r^{n}}\rho^{\otimes n}_{c,\ulines^{n}})
 \nonumber
\end{align*}
where $\rho^{\otimes n}_{c,\ulines^{n}} = \otimes_{i=1}^{n}\rho_{x_{1i}(s_{1}^{n})x_{2i}(s_{2}^{n})}$, where $e_{j}(s_{j}^{n}) = (x_{j1}(s_{j}^{n}),x_{j2}(s_{j}^{n}),\cdots, x_{jn}(s_{j}^{n}))$ for $j\in [2]$.
\end{definition}
\begin{definition}
 A function $f$ of the sources $\mathbb{W}_{S_{1}S_{2}}$ is said to be  reconstructable over a CQ-MAC
$\mathcal{N}_2$
 if for $\epsilon > 0$, $\exists$ a sequence $c_{f}^{(n)} = (n,e_{1}^{(n)},e_{2}^{(n)},\boldsymbol{\lambda})$ such that $\lim_{n \rightarrow \infty} \overline{\xi}(c_{f}^{(n)},\CalN_2) = 0$.
Restricting $f$ to a sum, we say the sum of sources $\mathbb{W}_{S_1S_2}$ over field $\FF_q$ is reconstructable over a CQ-MAC if $\CalS_1 = \CalS_2 = \FF_q$ and the function $f(S_1,S_2) = S_1 \oplus S_2$ is reconstructable over the CQ-MAC.
\end{definition}

We review the performance limit achievable using unstructured code ensembles in the following. 
\begin{prop} \label{prop:unstructured}
A function $f$ of the sources $\mathbb{W}_{S_{1}S_{2}}$ is   reconstructible over a CQ-MAC
$\mathcal{N}_2$ if
\begin{align}
    H(S_1,S_2) < \max_{p_{X_1}p_{X_2}} I(X_1X_2;Z)_{\sigma},
\end{align}
where the mutual information is defined for the following classical-quantum state
$$\sigma \deq \sum_{x_1x_2}p_{X_1}(x_1)p_{X_2}(x_2)\rho_{x_1x_2}\tensor \ketbra{x_1}\tensor\ketbra{x_2}.$$
\end{prop}
\begin{proof}
 The technique involves using the Slepian-Wolf \cite{slepian1973noiseless} source coding to compress the source to $H(S_1,S_2)$ bits, and followed by the Winter's channel coding over the CQ-MAC $\mathcal{N}_2$  \cite{winter2001capacity}.
\end{proof}

The objective of our work is to characterize improved sufficient conditions under which a generic bivariate function of the sources is reconstructible over a CQ-MAC $\mcl{N}_2$ by developing a framework that combines unstructured coding (as described in Proposition \ref{prop:unstructured}) and structured coding (as proposed in \cite{touheed_compCQMAC}) techniques for this problem.

\section{Main Results}
\label{sec:mainResults}
As an intermediate step toward providing the main result, we present an intermediary result that will be useful in obtaining the main result, and can also be of independent interest.
\subsection{{4-to-3 decoding over CQ-MAC}}
\label{subsec:4userCQMAC}
In this subsection, we consider the problem of 4-to-3 decoding over a
 $4$-user CQ-MAC, where the receiver aims to compute functions of messages of user 1 and 2, and the individual message of users 3 and 4. 
 Consider a (generic) 4-user CQ-MAC $\CalN_4$, which is specified through (i) finite (input) sets $\mathcal{V}_j \colon j \in [2] \mbox{ and } \mathcal{U}_j \colon j \in [2]$, (ii) a (output) Hilbert space $\CalH_Z$, and (iii) a collection of density operators $ (\rho_{v_{1}v_{2}u_{1}u_{2}} \in \mathcal{D}({\CalH_Z}): (v_{1},v_{2},u_{1},u_{2}) \in \mathcal{V}_{1} \times \mathcal{V}_{2} \times \mathcal{U}_{1} \times \mathcal{U}_{2})$. 

\begin{definition}
 \label{defn:4to3cqmaccode}
A code 
 $c = \left(n,\FF_q,e_{V_j} \colon j \in [2],
 e_{U_j}\colon j \in [2] ,\boldsymbol{\lambda}\right)$
 of block-length $n$,
 for 4-to-3 decoding over CQ-MAC $\CalN_4$ 
 consists of four encoding maps $e_{V_j} \colon \FF_q^{l} \rightarrow \mathcal{V}_{j}^{n} : j \in [2]$, $e_{U_j} : [q^{l_j}] \rightarrow \mathcal{U}_{j}^{n} : j \in [2]$, and a POVM  $\boldsymbol{\lambda}= \{ \lambda_{\{m^{\oplus},m_3,m_4\}} \in \mathcal{P}(\CalH_Z) : (m^{\oplus},m_3,m_4) \in \FF_q^{l} \times [q^{l_1}] \times [q^{l_2}]\}$, where $m^{\oplus} \define m_1 \oplus m_2$,
 $l$, $l_1$ and $l_2$ are positive integers, and $q$ is a prime number.
\end{definition}

\begin{definition}
Given a CQ-MAC $\mathcal{N}_4$, and a prime $q$, 
a rate triple $(R,R_1,R_2) > \ul{0}$ is said to be achievable for 4-to-3 decoding over the CQ-MAC 
if given any sequence of triples $(l(n),l_1(n),l_2(n))$, 
such that $ \limsup_{n\rightarrow \infty} \frac{l(n)}{n} \log q < R,
 \limsup_{n\rightarrow \infty} \frac{l_i(n)}{n} \log q < R_i\colon i \in [2]$, and any sequence $p_{M_{1}M_{2}M_{3}M_{4}}^{(n)}$ of PMFs on $\FF_q^{l} \times \FF_q^{l} \times [q^{l_1}] \times [q^{l_2}]$, 
there exists a code $c^{(n)} = (n,\FF_q,e_{V_j} \colon j \in [2], e_{U_j}\colon j \in [2] ,\boldsymbol\lambda)$  for 4-to-3 decoding over CQ-MAC $\mathcal{N}_4$  
of block-length $n$ 
such that 
\begin{align*}
    &\limsup_{n\rightarrow \infty}  \avgPeN =  \limsup_{n\rightarrow \infty}  1-  \sum_{\substack{\ulinem}} p_{\ulineM}(\ulinem)\tr(\lambda_{\{m^{\oplus},m_3,m_4\}}\rho_{\ulinem}^{\otimes n}) = 0,
\end{align*}
where $\rho^{\otimes n}_{\ulinem} \define \rho_{v_1^n(m_1)v_2^n(m_2)u_1^n(m_3)u_2^n(m_4)}$ $= \otimes_{i=1}^{n}\rho_{v_{1i}(m_{1})v_{2i}(m_{2})u_{1i}(m_{3})u_{2i}(m_{4})}$ (assuming $n$-independent uses of $\CalN_4$).
The convex hull of the union of the set of all achievable rate triples $(R,R_1,R_2)$ is the capacity region of the 4-to-3 decoding over CQ-MAC $\mathcal{N}_4$ and prime number $q$. 
\end{definition}

\begin{definition}
\label{defn:admpmfsandquantumstates}
Given a CQ-MAC $\CalN_4$ and a prime $q$, let $\mathscr{P}(\CalN_4,~q)$ be defined as collection of PMF
 $\{p_{\ulineV\ulineU}
\colon
 p_{\ulineV \ulineU} = p_{V_{1}}p_{V_{2}}p_{U_{1}}p_{U_{2}} \mbox{ is a PMF on }
\ul{\CalV} \times \ul{\CalU}
\}.$
For $p_{\ulineV\ulineU}
 \in \mathscr{P}(\CalN_4,q)$,
let $\mathscr{R}(p_{\ulineV\ulineU}
)$ be the set of rate triple $(R,R_1,R_2)$ such that the following inequalities holds:
\begin{align*}
    R &\leq I(V;Z|U_1,U_2)_{\sigma} - I_{\max}(V_1,V_2,V)_\sigma,\\
    R_1 &\leq I(U_1;Z|V,U_2)_{\sigma}, \\
    R_2 &\leq I(U_2;Z|V,U_1)_{\sigma}, \\
    R + R_1 &\leq I(V, U_1;Z|U_2)_{\sigma} - I_{\max}(V_1,V_2,V)_\sigma,\\
    R + R_2 &\leq I(V, U_2;Z|U_1)_{\sigma}- I_{\max}(V_1,V_2,V)_\sigma, \\
    R_1+R_2 &\leq I(U_1, U_2;Z|V)_{\sigma} \\
    R + R_1 + R_2 &\leq  I(V, U_1, U_2;Z) - I_{\max}(V_1,V_2,V)_{\sigma},
\end{align*}
where $I_{\max}(V_1,V_2,V)_{\sigma} = \max\{I(V_1; V)_{\sigma}, I(V_2;V)_{\sigma}\}, ~ V = V_1\oplus V_2$ and the mutual information quantities  are taken with respect to the classical-quantum state: \begin{align*}
\sigma \define &\sum_{\ulinev,\ulineu,v} p_{\ulineV\ulineU}(\ulinev,\ulineu)
\mathds{1}_{\{ v = v_{1}\oplus v_{2}\}} \ketbra{v}_{\mathrm{V}}
\otimes
 \ketbra{v_1}_{\mathrm{V_1}}  \otimes
 \ketbra{v_2}_{\mathrm{V_2}} \otimes
 \ketbra{u_1}_{\mathrm{U_1}}  \otimes
 \ketbra{u_2}_{\mathrm{U_2}}
 \otimes  \rho_{\ulinev\ulineu}.
\end{align*}  
Let $$ \mathscr{R}(\CalN_4,q) \define \displaystyle {\mathrm{c.c.}} ~\bigcup_{ \substack{p_{\ulineV\ulineU} 
\in \mathscr{P}(\CalN_4
,q)} }\mathscr{R}(p_{\ulineV\ulineU }
).$$ 
\end{definition}

\begin{theorem} \label{thm:4-to-3}If the rate triple $(R,R_1,R_2) \in \mathscr{R}(\CalN_4,q)$, then  $(R,R_1,R_2)$ is achievable for  4-to-3 decoding over a CQ-MAC $\CalN_4$ and prime $q$. 
\end{theorem}
\begin{proof} The proof is provided in Section$~$\ref{sec:proof4to3}.
\end{proof}

\subsection{Decoding arbitrary function f over CQ-MAC $\mathcal{N}_2$}
\label{subsec:sumOfSources}
 Here we provide our main result characterizing the sufficient conditions on the sources, for any reconstruction of the bivariate function $f$ at the decoder of the given CQ-MAC $\mcl{N}_2$.
Before we proceed, we provide the following definition for embedding a function into a finite field.


\begin{definition}
\label{def:embed}
A function $f:\underline{\CalS} \rightarrow \CalR$ of sources $\mathbb{W}_{S_{1}S_{2}}$ is said to be embeddable in a finite field $\fieldq$ if there exists (i) a pair of functions $h_j : \CalS_j \rightarrow \fieldq$ for $j\in [2]$, and (ii) a function $g: \fieldq \rightarrow \mcl{R},$  such that $\mathbb{W}_{S_1S_2}(f(S_1,S_2) = g(h_1(S_1)\oplus h_2(S_2))) = 1.$
\end{definition}
\begin{remark}
Note that for any given function $f$, the set of prime $q$ for which $f$ is embeddable with respect to $\FF_q$ is always non-empty. To see this, take $q > |\mcl{S}_1||\mcl{S}_2|,$ and let $h_1$ be any one-to-one mapping from $|\mcl{S}_1|$ to $\{0,1,...,|\mcl{S}_1|-1\}$, and let $h_2$ be any one-to-one from $|\mcl{S}_2|$ to $\{0,|\mcl{S}_1|,2|\mcl{S}_1|,...,|\mcl{S}_1|(|\mcl{S}_2|-1)\}$. Then, $h_1(\cdot)\oplus h_2(\cdot)$ is a one-to-one map from $|\mcl{S}_1|\times |\mcl{S}_2|$ to $\FF_q$ (see \cite[Def. 3.7]{pradhanalgebraic}). For example, the nonlinear logical OR ($\lor$) function of binary sources with $\mathcal{S}_1=\mathcal{S}_2=\{0,1\}$ can be embedded in $\mathbb{F}_3$, by noting that 
$S_1 \lor S_2=g(h_1(S_1) \oplus_3 h_2(S_2))$, where $g$ is given by $g: 0 \mapsto 0$, $1 \mapsto 1$, and  $2 \mapsto 1$, and $h_i$s are identity maps.
\end{remark}

\begin{definition}
\label{def:sumofsources}
Given the source $(\mathcal{S}_1,\mathcal{S}_2,\mathbb{W}_{S_1S_2},f)$, consider a prime $q$ such that $f$ is embeddable (according to Definition \ref{def:embed}) in $\fieldq$. Let $\CalP$ be the set of PMFs $p_{QW_1W_2|S_1S_2}$ defined on  $\CalQ\times \CalW_1 \times \CalW_2$ such that (a) $Q$ and $(S_1,S_2)$ are independent, (b) $W_1-S_1Q-S_2Q-W_2$ forms a Markov chain, and (c) $\CalQ, \CalW_1, \CalW_2$ are finite sets. For $p_{QW_1W_2|S_1S_2} \in \CalP$, let us define,
\begin{align*}
    &\mathscr{R}_S(p_{QW_1W_2|S_1S_2},q)  \define \Big\{ (R,R_1,R_2) \colon R \geq H(S|W_1W_2Q),  {R}_1 \geq I(S_1;W_1|QW_2), 
    {R}_2 \geq I(S_2;W_2|QW_1),
    \\ & \hspace{200pt} {R}_1 + {R}_2 \geq I(S_1S_2;W_1W_2|Q)\Big\},
\end{align*} where $S = h_1(S_1)\oplus h_2(S_2)$. Define 
 \begin{align*}
     \mathscr{R}_s(\mathbb{W}_{S_1S_2},f,q) &\define\mathrm{c.c.} \bigcup_{p\in \CalP} \mathscr{R}_S(p,q). 
 \end{align*}
\end{definition}
\begin{definition}
Given a CQ-MAC $\mathcal{N}_2$, and prime $q$, let $\mathscr{P}$ be the set of PMFs $p_{X_1|U_1V_1}$ and $p_{X_2|U_2V_2}$ with the input alphabets $(\CalU_1,\CalV_1)$ and $(\CalU_2,\CalV_2)$, and output alphabets $\CalX_1$ and $\CalX_2$,  respectively.
Define, 
\begin{align*}
    \mathscr{R}_C(p_{X_1|U_1V_1},p_{X_2|U_2V_2},q) = \mathscr{R}(\CalN_4,q),
\end{align*} where the corresponding $4$-user CQ-MAC $\mathcal{N}_4$ is characterized as: $$\rho_{\ulinev\ulineu} = \displaystyle \!\!\sum_{x_1x_2}\!\! p_{X_1|U_1V_1}(x_1|u_1v_1) p_{X_2|U_2V_2}(x_2|u_2v_2)\rho_{x_1x_2}.$$
Define,
\begin{align*}
     \mathscr{R}_c(\mathcal{N}_2,q) &\define \mathrm{c.c} \bigcup_{\big\{\substack{p_{X_j|U_jV_j} \colon j \in [2]} \big\} \in \mathscr{P}}\!\!\! \mathscr{R}_C(p_{X_j|U_jV_j}\colon j\in [2],q). 
 \end{align*}
\end{definition}

\begin{theorem}\label{thm:decodingArbitraryfunc}
If $\mathscr{R}_s(\mathbb{W}_{S_1S_2},f,q) \subset \mathscr{R}_c(\mathcal{N}_2,q)$ for some prime $q$, then the bivariate function $f$ of the sources $\mathbb{W}_{S_1S_2}$ is reconstructible over the CQ-MAC $\mathcal{N}_2$.
\end{theorem}
\begin{proof}
The proof is provided in Section \ref{sec:proofsum}.
\end{proof}


\subsection{Examples}
\label{subsec:examples}
We provide the following example to compare the sufficient conditions obtained in \cite[Theorem 2]{touheed_compCQMAC} and Theorem \ref{thm:decodingArbitraryfunc}. 
\begin{example}

Let $\mathcal{X}_1=\mathcal{X}_2=\mathcal{S}_1=\mathcal{S}_2=\mathcal{X}=\{0,1\}$, 
$\mathcal{H}_Z = \mathbb{C}^2$,
and 
$\rho_{x_1,x_2}=(1-\eta)\sigma_{x} + \eta \sigma_{\bar{x}}$, 
where $x \deq x_1 \oplus_2 x_2$ and  $\sigma_0,\sigma_1 \in \mathcal{D}(\mathcal{H}_Z)$ defined as \begin{align}
    \sigma_0 \deq \begin{pmatrix}
                    0.9545 & 0.0455i \\ 
                    -0.0455i & 0.0455
                \end{pmatrix}, \quad 
    \sigma_1 \deq \begin{pmatrix}
        0.0455 & 0.0455i \\
        -0.0455i  & 0.9545
    \end{pmatrix}. 
\end{align} 
Let $\rho(\eta) \define (1-\eta)\sigma_0+\eta\sigma_1$. 
Furthermore, to induce asymmetry in the rate region, we constrain one of the inputs with a cost constraint: $\mathbb{E}(X_1) \leq c$. We choose $c=0.1$ for the illustration. 
Let $S_1$ and $S_2$ be two highly asymmetric  correlated sources (as considered in \cite[Example 4]{ahlswede1983source}) with the following distribution:
\begin{align}
    &\mathbb{W}_{S_1,S_2}(0,0) = 0.003920,\; \mathbb{W}_{S_1,S_2}(0,1) = 0.976080,\; \mathbb{W}_{S_1,S_2}(1,0) = 0.019920, \; \mathbb{W}_{S_1,S_2}(1,1) = 0.000080, \nonumber 
\end{align}
and $P(S_1 = 0) = 0.98, P(S_2 = 0) = 0.023840$. 
Let $f(S_1,S_2)=S_1 \oplus_2 S_2$. Consider the sufficient conditions given by the unstructured coding scheme as provided in Proposition \ref{prop:unstructured}. 
 \begin{align}\label{eq:ex3_compunstruct}
     H(S_1,S_2) <
\max_{p_{X_1X_2}}\chi(\{p_{X_1X_2}(x_1,x_2),\rho_{x_1,x_2}\}), 
 \end{align}
with $X_1$ and $X_2$ being independent. Figure \ref{fig:subOpt} (left) depicts the behaviour of the right hand side of the above inequality for different values of $\eta$. We observe that the inequality fails for $\eta > 0.23$, and hence the function $f$ cannot be reconstructed using the technique based on unstructured codes (more formally, as described in Proposition \ref{prop:unstructured}) for all $\eta \in (0.23,0.5)$.
As for employing algebraic structured codes, in particular nested coset codes, as proposed in \cite{touheed_compCQMAC}, we observe the following. Using \cite[Theorem 2]{touheed_compCQMAC} and the fact that $H(S_1 \oplus S_2) = 0.0376223$, we obtain the sufficient conditions for the function $f$ to be reconstructible using only structured codes as
\begin{align}\label{eq:ex3_comp}
    0.0376223 <\max_{p_{X_1}p_{X_2}} \min\{H(X_1),H(X_2)\}-H(X) +
\chi(\{p_{X}(x),\rho_{x}\}),
\end{align}
where $X \deq X_1 \oplus_2 X_2$ and $\rho_x \deq \rho_{x_1x_2}$ for any $(x_1,x_2)$ such that $x = x_1 \oplus x_2$.
Figure \ref{fig:subOpt} (right) depicts the behaviour of the right hand side of the above inequality for different values of $\eta$. In particular, the inequality fails for $\eta > 0.11$, and as a result the function $f$ cannot be reconstructed using the approach of \cite[Theorem 2]{touheed_compCQMAC} for all $\eta \in (0.11,0.5)$. 
Now, we consider the sufficient conditions obtained from Theorem \ref{thm:decodingArbitraryfunc}. Figure \ref{fig:example3_intersection} shows the regions $\mathscr{R}_s(\mathbb{W}_{S_1S_2},f,2)$  and  $\mathscr{R}_c(\mathcal{N}_2,2)$ for different values of $\eta$, which demonstrates a clear overlap between $\mathscr{R}_s $ and $\mathscr{R}_c$ for $\eta = 0.20$ and $\eta = 0.25$, which implies that the function remains reconstructible for these $\eta$ values. 


\begin{figure}[htb]
    \centering
    \includegraphics[scale=0.55]{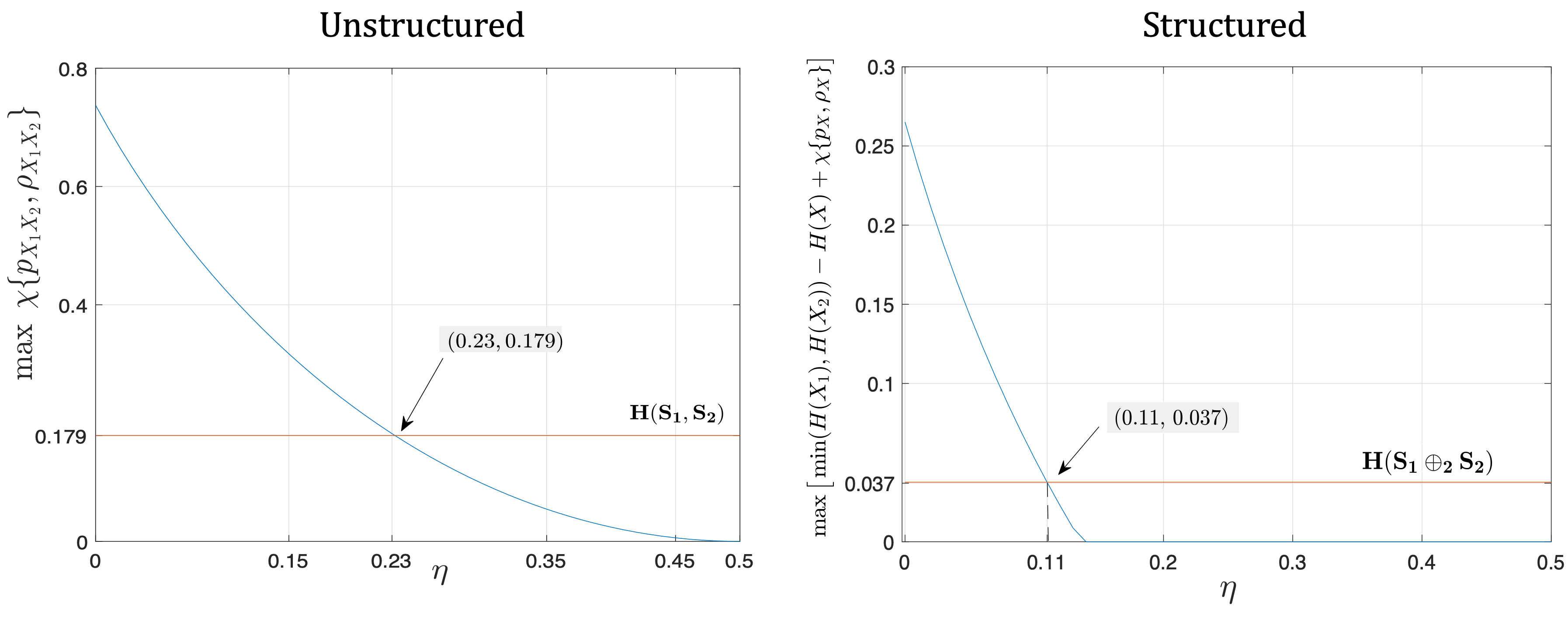}
    \caption{(Left) The variation of the right hand side of \eqref{eq:ex3_compunstruct}, and its intersection with $H(S_1 , S_2)$. This implies that the inequality in \eqref{eq:ex3_compunstruct} is not satisfied for all $\eta > 0.23$, and hence the function $f$ cannot be reconstructed using the approach based on unstructured coding (as described in Proposition \ref{prop:unstructured}). (Right) The variation of the right hand side of \eqref{eq:ex3_comp}, and its intersection with $H(S_1 \oplus S_2)$. This implies that the inequality in \eqref{eq:ex3_comp} is not satisfied for all $\eta > 0.11$, and hence the function $f$ cannot be reconstructed using the structured coding technique proposed in \cite{touheed_compCQMAC}.}
    \label{fig:subOpt}
\end{figure}

\begin{figure}[!htb]
    \centering
    \includegraphics[scale=0.55]{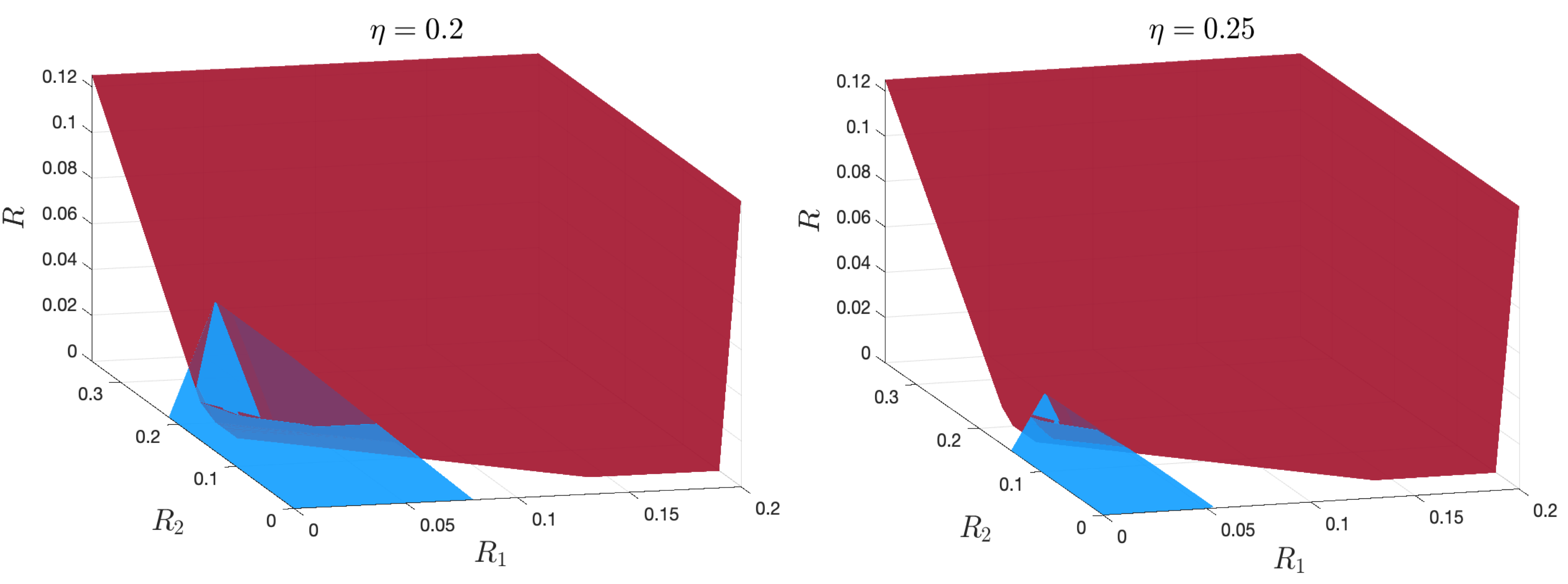}
    \caption{A depiction of the intersection of $\mathscr{R}_s$ (in red) and $\mathscr{R}_c$ (in blue) for  $\eta = 0.20$ and $\eta = 0.25$, using the framework based on both unstructured and structured coding, as described in Theorem \ref{thm:decodingArbitraryfunc}.}
    \label{fig:example3_intersection}
\end{figure}
\end{example}

\section{Proof of Theorem \ref{thm:4-to-3}}
\label{sec:proof4to3}
 Let $p_{\ulineV\ulineU}
 \in \mathscr{P}(\CalN_4,q)$ 
 be a PMF 
 $\mbox{ on }
\mathcal{V}_1 \times \mathcal{V}_2 \times \mathcal{U}_{1} \times \mathcal{U}_{2}$ where $\mathcal{V}_1 = \mathcal{V}_2 = \fieldq$. We begin by describing the coding scheme in terms of a specific class of codes. In order to choose codewords of a desired empirical distribution $p_{V_{j}}$, we employ Nested Coset Code (NCC), as described below.
\begin{definition}
 \label{def:NCC}
 An $(n,k,l,g_{I},g_{O/I},b^{n},e)$ NCC built over a finite field $\mathcal{V}=\fieldq$ comprises of (i) generator matrices $g_{I} \in \mathcal{V}^{k \times n}$, $g_{O/I} \in \mathcal{V}^{l \times n}$ (ii) a bias vector $b^{n}$, and (iii) an encoding map $e :\mathcal{V}^{l} \rightarrow \mathcal{V}^{k}$. We let $v^{n}(a,m) = ag_{I}\oplus_{q}mg_{O/I}\oplus_{q}b^{n}: (a,m) \in \mathcal{V}^{k} \times \mathcal{V}^{l}$,
for $a = e(m),$ and denoted as $a_m$.  
\end{definition}
Now, we intend to use the above mapping $v^{n}(a,m)$ from $\mathcal{V}^{l} \rightarrow \mathcal{V}^{n}$, as a part of the encoder in relation to Definition \ref{def:CQ-MACcode}.
Both the encoders $e_{V_j} \colon j \in [2]$ employ cosets of the \textit{same} linear code. We then consider a 4-to-3 decoding over a `perturbed' variant of CQ-MAC, which we denote as $\CalN_4'$. Note that in the current problem, the decoder wishes to decode three messages simultaneously and hence we use the framework of CQ joint typicality developed using the ideas of tilting, smoothing and augmentation \cite{sen2021unions}. This allows us to perform 
\textit{intersection of non-commuting POVM elements} to construct a set of POVMs for $\CalN_4'$. Finally, towards bounding the average error probability for $\CalN_4$, we use an argument,  similar to \cite[Equation 5]{sen2021unions}, which shows that the outputs of the channel $\CalN'_4$ and $\CalN_4$ are indistinguishable in trace  norm. Thus, the POVMs constructed for $\CalN_4'$ can be used for $\CalN_4$ with an additional boundable error term.



We now define a 4-to-3 decoding over `perturbed' CQ-MAC $\CalN_4'$ that consists of the following: (i) Finite (augmented input) sets $(\mathcal{V}_j \times \CalW_{V_j}), ~ (\mathcal{U}_j \times \CalW_{U_j}) \colon j \in [2]$. (ii) An (extended output) Hilbert space $$\CalH'_Z = \HZhat \bigoplus(\HZhat \otimes \CalW_{V_1}) \bigoplus(\HZhat \otimes \CalW_{V_2}) \bigoplus(\HZhat \otimes \CalW_{U_1}) \bigoplus(\HZhat \otimes \CalW_{U_2}),$$
where $\HZhat = (\CalH_Z \otimes\mathbb{C}^2)$, and   $\CalW_{V_j}$  denotes both a finite alphabet as well as a Hilbert space with dimension given by $|\CalW_{V_j}|$. The states in this Hilbert space are used as quantum registers to store classical values. 
Similarly $\CalW_{U_j}$
is defined. 
(iii) A collection of density operators
\[ \{\rho_{\ulinev\ulineu\ulinel}' \in \mathcal{D}({\CalH'_Z}): (\ulinev,\ulineu,\ulinel_V, \ulinel_U) \in \ul{\mathcal{V}} \times \ul{\mathcal{U}} \times\ul{\CalW}_{V} \times \ul{\CalW}_{U}\},\] where $\ulinel = (w_{V_1},w_{V_2},w_{U_1},w_{U_2}), ~\ulinel_V = (w_{V_1},w_{V_2}), \mbox{ and } \ul{\CalW}_{V} = \CalW_{V_1} \times \CalW_{V_2}$. Similarly $\ulinel_U \mbox{ and } \ul{\CalW}_{U}$ are defined.
Note that the states in the Hilbert spaces $\CalW_{V_j}$ and $\CalW_{U_j}$ are used as quantum registers to store classical values.
Define $\rho_{\ulinev\ulineu\ulinel}' \define \mathcal{T}_{\ulinel; \tau}^{\mathrm{V\ulineU}}(\tilde{\rho}_{\ul{v} \ul{u}} )$, where $\tilde{\rho}_{\ul{v} \ul{u}} \deq {\rho}_{\ul{v} \ul{u}} \otimes |0\ral\lal0|^{\CC^2}$, and $\mathcal{T}_{\ulinel; \tau}^\mathrm{V\ulineU}$ is a \textit{tilting map} \cite[Section 4]{sen2021unions}
from $\HZhat $ to $\CalH'_Z$  defined as: $$\mathcal{T}_{\ulinel; \tau}^\mathrm{V\ulineU}(\ket{z})  \define 
    \frac{(\ket{z} \bigoplus \tau\ket{z,w_{V_1}} \bigoplus  \tau\ket{z,w_{V_2}} \bigoplus \tau\ket{z,w_{U_1}} \bigoplus \tau\ket{z,w_{U_2}})}{\sqrt{1+4\tau^2}},$$ and $\tau$ will be chosen appropriately in the sequel.

    
 \par\noindent  \textbf{Encoding:}  
 Consider two NCCs $(n,k,l,g_{I},g_{O/I},b_j^{n},e_j)$
having the same parameters except with different bias vectors $b_j$s and encoding maps $e_j$s. 
 For each $j \in [2] \mbox{ and } m_{j} \in \fieldq^{l}$, let
 \begin{eqnarray}
\label{Eqn:NoTypicalElements}
 \mathcal{A}_{j}(m_{j}) \define \begin{cases} \{a_{m_{j}}: v^{n}_{j}(a_{m_{j}},m_{j}) \in T_{\delta}^{n}(p_{V_{j}})\}&\mbox{if }\theta(m_j) \geq 1 \\\{0^{k}\}&\mbox{otherwise,}\end{cases}
 \nonumber
\end{eqnarray}
where $\theta(m_j) \define \sum_{a \in \fieldq^{k}}\mathds{1}_{\left\{ v_j^{n}(a,m_j) \in T_{\delta}^{n}(p_{V}) \right\}}$.
 For $m_{j} \in \fieldq^{l}\colon j\in [2]$, a pre-determined element $a_{m_{j}} \in \CalA_{j}(m_{j})$ is chosen and let $v_{j}^{n}(a_{m_{j}},m_{j}) \define a_{m_{j}} {g_{I}} \;\oplus\; m_{j}g_{O/I}\;\oplus \;b_{j}^{n}$ for $ (a_{m_{j}},m_{j})\in \fieldq^{k+l} $ for $ j \in [2]$.
Moreover, for each $j\in [2]$ and $m_{j+2} \in [q^{l_j}]$, construct a codeword 
$u_{j}^{n}(m_{j+2}) \in \mathcal{U}_{j}^{n}$. Similarly, for each $j\in [2]$, $m_j \in \FF_q^l$ and $m_{j+2}\in [q^l_j]$, construct the codewords $w_{V_j}^n(m_j) \in \CalW_{V_j}^n$ and $w_{U_j}^n(m_{j+2}) \in \CalW_{U_j}^n$.
For later convenience, we define an additional 
identical map $w^n_V(m) = w^n_{V_1}(m)$ for all $m \in \FF_q^l$.
On receiving the message $\ulinem \in \fieldq^{l} \times \fieldq^{l} \times [q^{l_1}] \times [q^{l_2}]$, the quantum state 
$\rho_{\ulinem}'^{~\otimes n} \define$ 
 $$ \rho_{v_1^n(a_{m_1},m_1)w_{V_1}^n(m_1)v_2^n(a_{m_2},m_2)w_{V_2}^n(m_2)(u_1^n,w_{U_1}^n)(m_3)(u_2^n,w_{U_2}^n)(m_4)}' \label{encoding}$$
 is (distributively) prepared. 
Towards specifying a decoding POVM's, we define the following associated density operators.
\begin{align}
 \rho &\define \sum_{\ulinev^n, \ulineu^n} p_{\ulineV}^n(\ulinev^n) p_{\ulineU}^n(\ulineu^n)  \rho_{\ulinev^n \ulineu^n},  \label{def:statesforpovm1}\\
\rho_{v^n} &\define \sum_{\ulinev^n,\ulineu^n}p_{\ulineV|V}^n(\ulinev^n|v^n)
p_{\ulineU}^n(\ulineu^n)  \rho_{\ulinev^n \ulineu^n}, \nonumber\\ 
\rho_{u_i^n} &\define \sum_{\ulinev^n u_j^n}p_{U_j}^n(u_j^n)  \rho_{\ulinev^n \ulineu^n} \colon i\neq j, i,j \in [2], \nonumber\\
\rho_{v^n u_i^n} &\define \sum_{\ulinev^n, u_j^n} p_{\ulineV|V}^n(\ulinev^n|v^n) p^n_{U_j}(u_j^n) \rho_{\ulinev^n \ulineu^n} \colon i\neq j, i,j \in [2], \nonumber\\
&
\rho_{v^n\ulineu^n} = \sum_{\ulinev^n} p_{\ulineV|V}^n(\ulinev^n|v^n) \rho_{\ulinev^n\ulineu^n} \nonumber
\end{align}
$\mathrm{ where }~ p_{\ulineV|V}^n(\ulinev^n|v^n) \deq p_{\ulineV}^n(\ulinev^n) / p_{V}^n(v^n) \mathds{1}_{\{v_1^n\oplus v_2^n = v^n\}} .$
\par \noindent \textbf{Decoding:} The decoder is designed to decode the sum of the messages $m^{\oplus}$ along with the individual messages $m_3$ and $m_4$ transmitted over the `perturbed' 4-to-3 CQ-MAC $\CalN_4'$. To decode $m_3$ and $m_4$, we use the codebook used by the encoder, but to decode $m^\oplus$, we  
use the NCC $(n, k,l,g_I,g_{O/I},b^n,e)$, with all the parameters same as the NCCs used in the encoding, except that $b^n = b_1^n \oplus b_2^n$, and $e$ to be specified later.
Define $v^n(a,m) \deq ag_I + mg_{O/I} + b^n$.
representing a generic codeword and a generic coset, respectively.

\par \noindent\textbf{POVM construction} We start by defining the sub-POVMs
for channel $\CalN$, subsequently we will construct the sub-POVMs for the `perturbed' CQ-MAC $\CalN_4'$ using the process of \textit{tilting} \cite{sen2021unions}. Let $\pi_\rho$ be the typical projector for the state $\rho$.
Furthermore, for $j \in [2]$ and for all jointly typical vectors $(v^n,\ulineu^n)  \in \TDelta(p_{V\ulineU})$, 
let $ \pi_{v^n}, \pi_{u_j^n}, \pi_{v^n u_j^n}, \pi_{\ulineu^n}$ and $ \pi_{v^n\ulineu^n}$ be the conditional typical projector \cite[Def. 15.2.4]{2013Bk_Wil} with respect to the states $\rho_{v^n}, \rho_{u_i^n},\rho_{v^n u_i^n}, \rho_{\ulineu^n}$  and $\rho_{v^n\ulineu^n}$, respectively.
Now, we define the following sub-POVMs in the Hilbert space $\HZ$:
\begin{align}\label{sub-povm1}
\Piv \define \pi_\rho \pi_{v^n} \pi_{v^n\ulineu^n} \pi_{v^n} \pi_\rho, ~& 
\Piuj  \define \pi_\rho \pi_{u_j^n} \pi_{v^n\ulineu^n} \pi_{u_j^n} \pi_\rho, \nonumber \\
\Pivuj \define \pi_\rho \pi_{v^n u_j^n}  \pi_{v^n\ulineu^n} \pi_{v^n u_j^n} \pi_\rho, ~& \Piu  \define \pi_\rho \pi_{\ulineu^n} \pi_{v^n\ulineu^n} \pi_{\ulineu^n} \pi_\rho,
\nonumber \\
\Pivu \define \pi_\rho \pi_{v^n\ulineu^n} \pi_\rho  & \colon i\neq j,i,j \in [2].
\end{align}
The following is a well-known result regarding typical projectors and typical vectors 
$(\ulinev^n,\ulineu^n)  \in T_{8\delta}^{(n)}(p_{\ulineV\ulineU})$.

\begin{prop}\label{prop:povmproperty}
For all $\epsilon>0$, and $\delta \in (0,1)$ sufficiently small and $n$ sufficiently large, and $i,j \in [2]$ with $i \neq j$ the following inequality holds for the sub-POVMs defined in \eqref{sub-povm1}.
\begin{align*}
\tr\left (\Pi_{v^n\ulineu^n}^{\mathrm{\Phi}}\hspace{1pt} \rho_{\ulinev^n \ulineu^n} \right) &\geq 1-\epsilon, \hspace{10pt}\mbox{for all } \Phi \in\{ \mathrm{V},\mathrm{U_j},\mathrm{VU_j},\mathrm{\ulineU},\mathrm{V\ulineU} \}, \\
\tr\left (\Pivu \rho\right) &\leq 2^{-n(I(V, U_1, U_2;Z)_{\sigma}-\epsilon)}, \\
\sum_{\ulineu^n} p_{\ulineU}^n(\ulineu^n) \tr\left (\Piv \rho_{v^n} \right) &\leq 2^{-n(I(U_1,U_2;Z|V)_{\sigma}-\epsilon)},  \\
\sum_{v^n} p_{V}^n(v^n) \tr\left (\Piu \rho_{\ulineu^n} \right) &\leq 2^{-n(I(V;Z|U_1,U_2)_{\sigma}-\epsilon)}, \\
\sum_{u_i^n} p_{U_i}^n(u_i^n) \tr\left (\Pivuj \rho_{v^n u_j^n} \right) &\leq 2^{-n(I(U_i;Z|U_j,V)_{\sigma}-\epsilon)},  \\
\sum_{v^n u_i^n} p_{V}^n(v^n) p_{U_i}^n(u_i^n)  \tr\left (\Piuj \rho_{u_j^n} \right) &\leq 2^{-n(I(V, U_i;Z|U_j)_{\sigma}-\epsilon)}.
\end{align*}
\end{prop}

After constructing the sub-POVMs, we now construct the projectors. It is worth to observe that by the Gelfand-Naimark theorem \cite{holevo}, there exists orthogonal projectors $\projv, \projuj, \projvuj,\proju$ and $\projvu$ in $\HZhat^{\otimes n}$ that gives the same measurements statistics on the states $\big(\sigma \otimes {\ovec}\big) \in \CalD(\HZhat^{\otimes n})$ that sub-POVMs defined in \eqref{sub-povm1} give on the states $\sigma \in \CalD(\HZ)$. To summarize upto this point we have constructed the projectors in $\HZhat^{\otimes n}$ for the channel $\CalN_4$ using the sub-POVMs defined in \eqref{sub-povm1}, and we are now equipped to construct the sub-POVMs for $\CalN_4'$. 
Let us define $\spv$ as the orthogonal complement of the support of $\projv$. Analogously, we define $\spuj,\spvuj,\spu,$ and $\spvu$. Then we define the corresponding tilted subspace in $\HZprime$ as: $\tspv \define \mathcal{T}_{{w_V^n};\varepsilon}^{\mathrm{V}}(\spv)$
, for all $w_{V}^n \in \CalW_{V_1}^n$. Likewise, define $\tspuj, \tspvuj$ and $ \tspu$. 
Also, let us define a new subspace $\tspvu$, which is analogous to the `union' of `complement' of orthogonal projectors corresponding to the sub-POVMs defined in (\ref{sub-povm1}).
\begin{align}
    \tspvu \define 
\spvu \bigoplus \tspv  \bigoplus_{j\in [2]} \tspuj   \bigoplus_{j \in [2]} \tspvuj \bigoplus \tspu.
\end{align}
Consider a collection of orthogonal projectors $\widehat{\Pi}'_{v^n\ulineu^n w_V^n \ulinew_U^n}$ in $\HZprime$ projecting onto $\tspvu$, and the orthogonal projector $\widetilde\Pi'$  projecting onto  $\HZhat^{\otimes n}$.
Subsequently, define the sub-POVMs in $\HZprime$ for channel $\CalN'_4$ as follows:
$$\gamma_{v^n \ulineu^n w_V^n \ulinel_U^n} \define \left(I - \widehat{\Pi}'_{v^n\ulineu^n w_V^n \ulinew_U^n}\right) ~ \widetilde\Pi' ~\left(I - \widehat{\Pi}'_{v^n\ulineu^n w_V^n \ulinew_U^n}\right),$$
The decoder now uses the sub-POVMs $\gamma_{v^n \ulineu^n w^n_V \ulinel_U^n}$ as defined above, to construct a \textit{square root measurement} \cite{holevo,2013Bk_Wil} to decode the messages, we define following operators,
\begin{align}\holevoPOVM \define {\Big(\sum_{\hat{a},\hat{m}}  \sum_{\hat{m}_3,\hat{m}_4}\gamma_{(\hat{a},\hat{m}),\hat{m}_3,\hat{m}_4}\Big)}^{-1/2}\ \gamma_{(a,m),m_3,m_4} ~ {\Big(\sum_{\hat{a},\hat{m}}  \sum_{\hat{m}_3,\hat{m}_4}\gamma_{(\hat{a},\hat{m}),\hat{m}_3,\hat{m}_4}\Big)}^{-1/2}, \end{align}
where $\gamma_{(a,m),m_3,m_4}$ is an abbreviation for $\gamma_{v^n(a,m) u_1^n(m_3)u_2^n(m_4)w_V^n(m)w_{U_1}^n(m_3)w_{U_2}^n(m_4)}$, and we let the perturbation $w_V$ used by the decoder is identical to that of either 
user 1 or user 2, and  without loss of generality  $w_{V} = w_{V_1}$, as mentioned earlier (in the discussion on encoding).

\textbf{Distribution of Random Code:} The distribution of the  random code is completely specified through the distribution $\CalP(\cdot)$ of $G_{I},G_{O/I}, B_{j}^{n}, A_{m_{j}}, , W_{V_j}^{n}(m_{j}), U_{j}^{n}(m_{j+2}),
W_{U_j}^{n}(m_{j+2}) \colon j \in [2]$. We let

\begin{align}
  \mathcal{P}\left(\!\!\!\!
  \begin{array}{c} G_{I}=g_{I},G_{O/I}=g_{O/I}, B_{j}^{n} = b_{j}^{n} , A_{m_{j}}=a_{m_{j}},\\
U_{j}^n(m_{j+2}) = u_{j}^{n}(m_{j+2}),   \\ W_{V_j}^n(m_{j}) = w_{v_j}^{n}(m_{j}),  W_{U_j}^n(m_{j+2}) = w_{u_j}^{n}(m_{j+2})\\  \colon j \in [2] \mbox{ , } \ulinem  \in \fieldq^{l} \times \fieldq^{l} \times [q^{l_1}] \times [q^{l_2}]   \end{array} \!\!\!\! \right) \!=\!\displaystyle \prod_{\substack{j \in [2]}}\frac{\mathds{1}_{\{a_{m_j} \in \CalA_{j}(m_{j}) \}}}{\theta(m_{j})|\CalW_{V_j}||\CalW_{U_j}|} 
\label{Eqn:DistOfRandomCode}
  \frac{p_{U_{j}}^n(u_{j}^{n}(m_{j+2}))}{q^{kn+ln+2n}}.  
\end{align}

\noindent \textbf{Error Analysis}: We derive an upper bound on $\avgPeNprime$, by averaging over the above ensemble. 
Our key insight for the error analysis will be similar to the those adopted in proof of \cite[Theorem 2]{touheed_compCQMAC} and \cite[Section 4]{sen2021unions}. Using the encoding and decoding rule stated above, the average probability of error of the code is given as,
\begin{align}
    \avgPeNprime 
    &= \sum_{\ulinem} p_{\ulineM}(\ulinem) \text{Tr}\left\{\left(I - \sum_{a} \lambda_{(a,m^\oplus),m_3,m_4} \right) \rho_{\ulinem}'^{~\otimes n}\right\}\nonumber \\
    & \leq \sum_{\ulinem} p_{\ulineM}(\ulinem) \text{Tr}\left\{\left(I - \lambda_{(a^{\oplus},m^{\oplus}),m_3,m_4} \right) \rho_{\ulinem}'^{~\otimes n}\right\}  \nonumber
\end{align}
where $a^{\oplus} \define a_{m_1} \oplus a_{m_2}$ and $\rho_{\ulinem}'$ is as defined in the Encoding section (Sec.~\ref{encoding}). 
Now consider the event, 
\begin{align}
 \mathscr{E} \define \left\{ \left( \begin{array}{c} V_{1}^{n}(A_{m_{1}},m_{1}), V_{2}^{n}(A_{m_{2}},m_{2}),  \\U_{1}^{n}(m_{3}),U_{2}^{n}(m_{4}), V^n(A^{\oplus},m^{\oplus}) \end{array} \right) \in T_{8\delta}^{(n)}(p_{\ulineV\ulineU V})  \right\} \nonumber
\end{align}
where $ V^n(A^{\oplus},m^{\oplus}) \deq V_{1}^{n}(A_{m_{1}},m_{1}) \oplus V_{2}^{n}(A_{m_{2}},m_{2})$.
Then, 
\begin{align*}
 \mathbb{E}_{\mathcal{P}}\left\{ \avgPeNprime \right\} &=
 \mathbb{E}_{\mathcal{P}}\left\{ \avgPeNprime \mathds{1}_{\mathscr{E}^{c}}
  + \avgPeNprime \mathds{1}_{\mathscr{E}}\right\}\leq
 \underbrace{\mathbb{E}_{\mathcal{P}}\left\{ \mathds{1}_{\mathscr{E}^{c}}\right\}}_{T_{1}}
  + \underbrace{\mathbb{E}_{\mathcal{P}}\left\{\avgPeNprime \mathds{1}_{\mathscr{E}}\right\}}_{T_{2}}.\nonumber
\end{align*}
 To bound the error $T_1$, we provide the following proposition.
\begin{prop}\label{prop:boundencerror} For all $\epsilon \in (0,1)$, and for all sufficiently large $n$ and sufficiently small $\delta$,  we have $\mathbb{E}_{\mathcal{P}}\left\{ \mathds{1}_{\mathscr{E}^{c}}\right\} \leq \epsilon $, if $ \frac{k}{n}\log{q} \geq \log{q} - \min\{H(V_1),H(V_2)\} + \delta$. 
\end{prop}
\begin{proof}
Refer to \cite[Appendix B]{padakandla2013computing} for the proof.
\end{proof}

To bound the error probability corresponding to $T_2$, we apply the Hayashi-Nagaoka inequality and obtain $$T_2 \leq \mathbb{E}_{\mathcal{P}}\Big[ 2\; T_{20} + 4 \underbrace{\Big\{T_{2V} + \sum_{j\in [2]} T_{2U_{j}} + \sum_{j \in [2]} T_{2VU_{j}}  + T_{2\ulineU} + T_{2V\ulineU} \Big\}}_{T_{21}}\Big],$$ 
where,
\begin{align}
\label{eqn:erroreventsafterhayashinagaol}
T_{20} &\define 1- \sum_{\ulinem}p_{\ulineM}(\ulinem) \tr(\Gamma_{(A^{\oplus},m^{\oplus}),m_3,m_4}\rho_{\ulinem}'^{~\otimes n} )\mathds{1}_{\mathscr{E}}, \nonumber 
\\
 T_{2U_{1}} &\define \sum_{\ulinem}\sum_{\hat{m}_{3} \neq m_{3}}p_{\ulineM}(\ulinem) \tr(\Gamma_{(A^{\oplus},m^{\oplus}),\hat{m}_3,m_4}\rho_{\ulinem}'^{~\otimes n} )\mathds{1}_{\mathscr{E}}, 
 \nonumber\\
 T_{2V} &\define  \sum_{\ulinem}\sum_{\hat{a}} \sum_{\hat{m} \neq m^\oplus}p_{\ulineM}(\ulinem) \tr(\Gamma_{(\hat{a},\hat{m}),m_3,m_4}\rho_{\ulinem}'^{~\otimes n} )\mathds{1}_{\mathscr{E}},
 \nonumber\\
 T_{2VU_1} &\define  \sum_{\ulinem}\sum_{\hat{a}} \sum_{\substack{\hat{m} \neq m^\oplus\\ \hat{m}_3 \neq m_3}} p_{\ulineM}(\ulinem) \tr(\Gamma_{(\hat{a},\hat{m}),\hat{m}_3,m_4}\rho_{\ulinem}'^{~\otimes n})\mathds{1}_{\mathscr{E}},
 \nonumber\\
 T_{2\ulineU} &\define \sum_{\ulinem} \sum_{\substack{\hat{m}_{3}\neq m_3 \\ \hat{m}_{4} \neq m_4}} p_{\ulineM}(\ulinem) \tr(\Gamma_{(A^{\oplus},m^{\oplus}),\hat{m}_3,\hat{m}_4}\rho_{\ulinem}'^{~\otimes n})\mathds{1}_{\mathscr{E}},
 \nonumber\\
 T_{2V\ulineU} &\define  \sum_{\ulinem}\sum_{\hat{a}} \sum_{\substack{\hat{m} \neq m^\oplus\\ \hat{m}_3 \neq m_3\\\hat{m}_4 \neq m_4}} p_{\ulineM}(\ulinem) \tr(\Gamma_{(\hat{a},\hat{m}),\hat{m}_3,\hat{m}_4} \rho_{\ulinem}'^{~\otimes n} )\mathds{1}_{\mathscr{E}}, \nonumber
\end{align} 
$\Gamma_{(A^{\oplus},m^{\oplus}),\hat{m}_3,\hat{m}_4}$ is a randomized version of  $\gamma_{(a^\oplus,m^\oplus),\hat{m}_3,\hat{m}_4}$. Similary we can define $T_{2U_2}$ and $T_{2VU_2}$.
Below, we provide the following propositions that summarize all the rate constraints obtained from bounding these error terms.
\begin{prop}\label{prop:erroranalysisT20}
For any $\epsilon \in (0,1) $, and for all sufficiently small $\delta, \tau > 0$ and sufficiently  large $n$,  
we have $\EE_{\mcl{P}}[T_{20}] \leq \epsilon$. 
\end{prop}
\begin{proof}
The proof is provided in Appendix \ref{appx:T20_proof}.
\end{proof}

\begin{prop}\label{prop:erroranalysis}
For any $\epsilon \in (0,1)$, and for  all 
sufficiently small $\delta,\tau > 0$ and
sufficiently large $n$, 
we have 
$\EE_{\CalP}[T_{21}] \leq \epsilon$ if the following inequalities hold:
 \begin{align*}
    \frac{2k+l_j}{n}\log q &\leq 2\log{q} + I( U_j;Z|V,U_i)_{\sigma} - H(V_1,V_2) -\epsilon, \\
    \frac{2k+l_1+l_2}{n}  \log q &  \leq 2\log{q}+I( U_1,U_2;Z|V)_{\sigma} - H(V_1,V_2) -\epsilon, \\
    \frac{3k+l}{n}\log q  &\leq  3\log q + I(V;Z|U_1,U_2)_{\sigma} - H_{V_1,V_2}-\epsilon, \\
    \frac{3k+l+l_j}{n} \log q &\leq 3\log q +  I(V, U_j;Z|U_i)_{\sigma} - 
    H_{V_1,V_2}-\epsilon,\\
    \frac{3k+l+l_1+l_2}{n} \log q & \leq 3\log q +  I(V, U_1,U_2;Z)_{\sigma} - H_{V_1,V_2} -\epsilon  ,
\end{align*}
 where  $ i,j \in [2], i \neq j, H_{V_1,V_2} = H(V_1,V_2) + H(V),\text{ and }$ the mutual information quantities are taken with respect to the classical-quantum state $\sigma$ same as in Definition \ref{defn:admpmfsandquantumstates}.
\end{prop}
\begin{proof}
The proof is provided in Appendix \ref{appx:proofPropErrorAnalysis}.
\end{proof}

Now, we need to bound average error probability for $\mathcal{N}_4$. 
For any $\epsilon \in (0,1)$, if we let $\tau  = \epsilon^{1/4}$, and  use the following argument, $\norm{\rho_{\ulinev^n\ulineu^n\ulinew^n}' - \tilde{\rho}_{\ulinev^n\ulineu^n}}_1 \leq 4\tau$ (similar to the provided in \cite[Equation 5]{sen2021unions}) and the trace inequality $\tr{\Delta\rho}\leq\tr{\Delta\sigma} + \frac{1}{2}\norm{\rho-\sigma}_1$, where $0\leq \Delta,\rho,\sigma \leq I$, then for all sufficiently large $n$, we have $\avgPeN  \leq  \avgPeNprime + 2\epsilon^{1/4}$. In other words, the average decoding error for CQ-MAC $\CalN_4$ is bounded from above by the average decoding error for CQ-MAC $\CalN_4'$  with an additional error of $2\epsilon^{1/4}$ for the same rate constraints and decoding strategy used for $\CalN'_4$.
This concludes the proof of Theorem$~\ref{thm:4-to-3}$.

\section{Proof of Theorem \ref{thm:decodingArbitraryfunc}}
\label{sec:proofsum}


We use the approach of source channel separation with two modules.
Consider a source given by $(\mathbb{W}_{S_1,S_2},f)$. For the source part, the theorem requires showing the above source can be compressed to rates $(R,R_1,R_2)$ that belongs to $\mathscr{R}_s(\mathbb{W}_{S_1,S_2},f,q)$. Ahlswede-Han \cite{ahlswede1983source} source coding scheme achieves this. 
This forms the source coding module. 
This module produces messages $M_{j1},M_{j2}$ at encoder $j \in[2]$, at rates $R_j,R$, respectively.
As for the channel part, its task is to recover $(M_{11},M_{21},M_{12}\oplus M_{22})$ reliably, and  to provide it to the source decoder. For this, we employ the result from Theorem \ref{thm:4-to-3}, which shows that if the triple $(R,R_1,R_2)$ belongs to $\mathscr{R}_c(\CalN_2,q)$, then for any arbitrary distribution of $p_{M_{11}M_{12}M_{21}M_{22}}$, such a recovery is guaranteed. This completes the proof of the theorem \ref{thm:decodingArbitraryfunc}.

\appendices

\section{Proof of Propositions}
\subsection{Proof of Proposition \ref{prop:erroranalysisT20} } 
\label{appx:proofPropAnalysisT20}
\label{appx:T20_proof}
\par
For $\ulinem$,  $a_{m_1}$ and $a_{m_2}$, define the following events:
\[\mcl{V}  \triangleq {\{ V_j^n(a_{m_j},m_j) = v_j^n \ : \ j \in [2]\}},
\quad \mcl{U} \triangleq {\{
U_j^n(m_{j+2}) = u_j^n \ : \ j \in [2]
\}}, \]
\[\mcl{W} \triangleq \{
\ulineW^n(\ulinem) = \ulinew^n\},
\quad 
\mcl{A} \triangleq \{A_{m_j} = a_j\ : \ j \in [2]\}.\]

Additionally, for $m^\oplus$ and ${a^\oplus}$, define the following events:
\[\h{\CalV} \define \{{V}^n({a^\oplus},{m^\oplus}) = {v}^n\}, \, \hat{\mcl{W}} \define \{
{W}_{V}^n({m^\oplus}) = {w}_{V}^n\}.\]
\begin{align*}
    & \EE_{\mathcal{P}}[T_{20}]
    = \mathbb{E}_\CalP \left[ \sum_{\substack{\ulinem\\\ulinea}} 
    \sum_{\substack{\ul{v}^n \ul{u}^n v^n\\\ulinew^n w_V^n}}
    p_{\ul{M}}(\ulinem)
    \tr\left\{
    \left(\!I\!-
    \Gamma_{v^n \ulineu^n w_V^n \ulinew_U^n}
    \right)
    \rho'_{\ul{u}^n 
    \ul{v}^n
    \ulinel^n}
    \right\}
    \bbm{1}_{\mcl{V}}
    \bbm{1}_{\h{\mcl{V}}}
    \bbm{1}_{\mcl{U}}
    \bbm{1}_{\mcl{W}}
    \bbm{1}_{\h{\mcl{W}}}
    \bbm{1}_{\mcl{A}}
    \bbm{1}_{\mcl{E}}
    \right], \\
    &\overset{(a)}{\leq}2\tau + \sum_{\substack{\ulinem\\\ulinea}} 
    \sum_{\substack{(\ul{v}^n \ul{u}^n) \in T_{8\delta}^{(n)}\\v^n \ulinew^n w_V^n}}
    p_{\ul{M}}(\ulinem)
    \tr\left\{
    \left(\!I\!-
    \Gamma_{v^n \ulineu^n w_V^n \ulinew_U^n}\right)
    \tilde\rho_{\ul{u}^n 
    \ul{v}^n}
    \right\}
    \mcl{P}(\mcl{V},\h{\CalV}, \mcl{A})
    \mcl{P}\left(\mcl{U}\right)
    \mcl{P}(\mcl{W,\h{W}}), 
    \\
    &\overset{(b)}{\leq} 2\tau + 4 \sum_{\substack{\ulinem\\\ulinea}} 
    \sum_{\substack{(\ul{v}^n \ul{u}^n) \in T_{8\delta}^{(n)}\\v^n \ulinew^n w_V^n}}
    p_{\ul{M}}(\ulinem)
    \left[\tr\left\{
    (\!I\!-
    \widetilde{\Pi}')
    \tilde\rho_{\ul{u}^n 
    \ul{v}^n}
    \right\} \right. \nonumber \\
    & \hspace{2in} \left. + \tr\left\{
    \widehat{\Pi}'_{v^n\ulineu^n w_V^n \ulinew_U^n}
    \tilde{\rho}_{\ul{u}^n 
    \ul{v}^n}
    \right\}
    \right] 
    \mcl{P}(\mcl{V},\h{\CalV}, \mcl{A})
    \mcl{P}\left(\mcl{U}\right)
    \mcl{P}(\mcl{W,\h{W}}),
    \\
    &\overset{(c)}{=} 2\tau + 4 \sum_{\substack{\ulinem\\\ulinea}} 
    \sum_{\substack{(\ul{v}^n \ul{u}^n) \in T_{8\delta}^{(n)}\\v^n \ulinew^n w_V^n}}
    {p_{\ul{M}}(\ulinem)} \11_{\{v^n = v_1^n \oplus v_2^n\}}
    \tr\left\{
    \widehat{\Pi}'_{v^n\ulineu^n w_V^n \ulinew_U^n}
    \tilde\rho_{\ul{u}^n 
    \ul{v}^n}
    \right\}
    \mcl{P}(\mcl{V},\h{\CalV}, \mcl{A})
    \mcl{P}\left(\mcl{U}\right)
    \mcl{P}(\mcl{W,\h{W}}),
    \\
    &\overset{(d)}{\leq} 2\tau  +
    4\cdot \frac{18}{\tau^2} \sum_{\substack{\ulinem\\\ulinea}}
    \sum_{\substack{(\ul{v}^n \ul{u}^n, v^n) \in T_{8\delta}^{(n)}}}
    {p_{\ul{M}}(\ulinem)}
    \left(~\sum_{\Phi \in \mathscr{U}  }\left(1-\tr\left\{
    \Pi^{\mathrm{\Phi}}_{v^n \ulineu^n}
   {\rho}_{\ul{v}^n \ul{u}^n}
    \right\}\right)
    \right)
    \mcl{P}(\mcl{V},\h{\CalV}, \mcl{A})
    \mcl{P}\left(\mcl{U}\right),\\ 
    &\overset{(e)}{\leq}  
    2\tau  + \frac{28\!\cdot\!18}{\tau^2}\delta_2,
\end{align*}

where $\mathscr{U} \deq \{\mathrm{V},\mathrm{U_j},\mathrm{VU_j},\mathrm{\ulineU},\mathrm{V\ulineU}\}$, and $\delta_2(\delta) \searrow 0 $ as $\delta \searrow 0$, and $(a)$ follows from the argument $\norm{\rho_{\ulinev^n\ulineu^n\ulinew^n}'-\tilde{\rho}_{\ulinev^n\ulineu^n}}_1 \leq 4\tau$ (similar to the \cite[Equation 5]{sen2021unions}) and the trace inequality $\tr{\Delta\rho}\leq\tr{\Delta\sigma} + \frac{1}{2}\norm{\rho-\sigma}_1$, where $0\leq \Delta,\rho,\sigma \leq I$
,  $(b)$ follows from Non-Commutative union bound \cite{sen2012achieving}, 
$(c)$ follows from the fact that $\widetilde{\Pi}'$ is a projection operator $\HZprime$ projecting onto  $\HZhat^{\otimes n}$, and $\tilde{\rho}_{\ulinev^n\ulineu^n} \in \CalD(\HZhat^{\otimes n})$. Thus, $\tr\{
    (I-\widetilde{\Pi}')
    \tilde\rho_{\ul{v}^n 
    \ul{u}^n}
    \} = 0$, $(d)$ follows from \cite[Corollary 1]{sen2021unions}, 
    and $(e)$ follows from Proposition \ref{prop:povmproperty}. Letting $\tau = \delta_2^{1/4}$, we obtain
    $\EE_{\CalP}[T_{20}] \leq 504 \sqrt{\delta_2} + 2 \delta_2^{1/4}.$
    
This concludes the proof of the Proposition \ref{prop:erroranalysisT20}.

\subsection{Proof of Proposition \ref{prop:erroranalysis}}
\label{appx:proofPropErrorAnalysis}

For $\ulinem$,  $a_{m_1}$ and $a_{m_2}$, define the following events:
\[\mcl{V}  \triangleq {\{ V_j^n(a_{m_j},m_j) = v_j^n \ : \ j \in [2]\}},
\quad \mcl{U} \triangleq {\{
U_j^n(m_{j+2}) = u_j^n \ : \ j \in [2]
\}}, \]
\[\mcl{W} \triangleq \{
\ulineW^n(\ulinem) = \ulinew^n\},
\quad 
\mcl{A} \triangleq \{A_{m_j} = a_j\ : \ j \in [2]\}.\]

\subsubsection{Analysis of $T_{2V\ulineU}$}
We begin by analyzing error event $T_{2V\ulineU}$
Define the following additional events for 
 $\hat{m},\hat{m}_3,\hat{m}_4$ and $\hat{a}$:
\[ \h{\CalV} \define \{\h{V}^n(\h{a},\h{m}) = \h{v}^n\},
\quad \h{\mcl{U}} \triangleq {\{
U_j^n(\h{m}_{j+2}) = \h{u}_j^n \ : \ j \in [2]
\}},\]
\[\hat{\mcl{W}} \deq \{
{W}_{V}^n(\h{m}) = \hat{w}_{V}^n,~{\ulineW}_{U}^n(\h{m}_3,\h{m}_4) = \hat{\ulinew}_{U}^n\}.\]

\begin{align*}
    \!\!&\mathbb{E}_{\mathcal{P}}\Big[T_{2V\ulineU}\Big]= \mathbb{E} \Bigg[ \sum_{\substack{\ul{m} \\ \ul{a}}} 
    \sum_{\substack{\ul{u}^n\h{\ul{u}}^n \ul{v}^n \h{v}^n\\ \ulinew^n \h{w}_{V}^n 
    \h{\ulinew}_{U}^n}} 
    \sum_{\h{a}}
    \sum_{\substack{\h{m} \neq m^\oplus \\ \h{m}_3 \neq m_3 \\ \h{m}_4 \neq m_4}}
    p_{\ul{M}}(\ul{m})
    \text{Tr}\Big\{
    \Gamma_{\h{v}^n \h{\ul{u}}^n \h{w}_{V}^n 
    \h{\ulinew}_{U}^n}
    \rho'_{\ul{v}^n \ul{u}^n \ulinew^n}
    \Big\}
    \bbm{1}_{\mcl{V}}
    \bbm{1}_{\h{\mcl{V}}}
    \bbm{1}_{\mcl{U}}
    \bbm{1}_{\h{\mcl{U}}}
    \bbm{1}_{\mcl{W}}
    \bbm{1}_{\h{\mcl{W}}}
    \bbm{1}_{\mcl{A}} 
    \bbm{1}_{\mcl{E}} 
    \Bigg],  \\
    &\overset{(a)}{\leq} 
    \sum_{\substack{\ul{m} \\ \ul{a}}}
    \sum_{\substack{(\ul{u}^n,  \ul{v}^n) \in T_{8\delta}^{(n)}\\ \h{\ul{v}}^n \h{u}^n\\ \ulinew^n\h{w}_{V}^n 
    \h{\ulinew}_{U}^n}} 
    \sum_{\h{a}}
    \sum_{\substack{\h{m} \neq m^\oplus \\ \h{m}_3 \neq m_3 \\ \h{m}_4 \neq m_4}}
    p_{\ul{M}}(\ul{m})
    \text{Tr}\Big\{
    \Gamma_{\h{v}^n \h{\ul{u}}^n \h{w}_{V}^n 
    \h{\ulinew}_{U}^n}
    \rho'_{\ul{v}^n \ul{u}^n
    \ulinew^n}
    \Big\}
    \mcl{P}(\mcl{V},\h{\mcl{V}})
    \mcl{P}(\mcl{U})
    \mcl{P}(\h{\mcl{U}})
    \mcl{P}(\mcl{W},\h{\mcl{W}}), \\
    &\overset{(b)}{=} 
    \sum_{\substack{\ul{m} \\ \ul{a}}}
    \sum_{\substack{\h{v}^n\h{\ul{u}}^n\\ \h{w}_{V}^n 
    \h{\ulinew}_{U}^n}} 
    \sum_{\h{a}}
    \sum_{\substack{\h{m} \neq m^\oplus \\ \h{m}_3 \neq m_3 \\ \h{m}_4 \neq m_4}}
    \frac{p_{\ul{M}}(\ul{m})}{q^{3n}}
    \mcl{P}(\h{\mcl{U}})
    \mcl{P}(\h{\mcl{W}})
    \text{Tr}\Big\{
    \Gamma_{\h{v}^n \h{\ul{u}}^n \h{w}_{V}^n 
    \h{\ulinew}_{U}^n}
    \sum_{\substack{(\ul{u}^n,  \ul{v}^n) \in T_{8\delta}^{(n)}\!\! \\ \ul{w}^n}} 
    \mcl{P}(\mcl{U})
    \mcl{P}(\mcl{W}|\h{\mcl{W}})
    \rho'_{\ul{v}^n \ul{u}^n
    \ul{w}^n}
    \Big\}, \\
    &\overset{(c)}{=}
    \sum_{\substack{\ul{m} \\ \ul{a}}} 
    \sum_{\substack{\h{v}^n\h{\ul{u}}^n\\ \h{w}_{V}^n 
    \h{\ulinew}_{U}^n}} 
    \sum_{\h{a}}
    \sum_{\substack{\h{m} \neq m^\oplus \\ \h{m}_3 \neq m_3 \\ \h{m}_4 \neq m_4}}
    \frac{p_{\ul{M}}(\ul{m})}{q^{3n}}
    \mcl{P}(\h{\mcl{U}})
    \mcl{P}(\h{\mcl{W}})\text{Tr}\Big\{
    \Gamma_{\h{v}^n \h{\ul{u}}^n \h{w}_{V}^n 
    \h{\ulinew}_{U}^n} \!\!\!
    \sum_{\substack{(\ul{u}^n,  \ul{v}^n) \in T_{8\delta}^{(n)}\!\!}}\!\!\!
    \mcl{P}(\mcl{U})
    \\
    & \hspace{20pt}\times 
    \Big[
    \11_{\{\hat{m} \neq m_1\}}
    \sum_{\ulinew^n} \frac{1}{|\mcl{W}|^{4n}}
    \mathcal{T}_{\ulinew^n; \tau}^\mathrm{\ulineV\ulineU}(\tilde{\rho}_{\ul{v}^n \ul{u}^n})
    +
    \11_{\{\hat{m} = m_1\}}
    \sum_{\ulinew^n} 
    \11_{\{w_{V_1}^n = \hat{w}_{V}^n\}}
    \frac{1}{|\mcl{W}|^{3n}}
    \mathcal{T}_{\ulinew^n; \tau}^\mathrm{\ulineV\ulineU}(\tilde{\rho}_{\ul{v}^n \ul{u}^n})
    \Big]
    \Big\}, \\
    &\overset{(d)}{\leq} \sum_{\substack{\ul{m} \\ \ul{a}}} 
    \sum_{\substack{\h{v}^n\h{\ul{u}}^n\\ \h{w}_{V}^n 
    \h{\ulinew}_{U}^n}} 
    \sum_{\h{a}}
    \sum_{\substack{\h{m} \neq m^\oplus \\ \h{m}_3 \neq m_3 \\ \h{m}_4 \neq m_4}}
    \frac{p_{\ul{M}}(\ul{m})}{q^{3n}}
    \mcl{P}(\h{\mcl{U}})
    \mcl{P}(\h{\mcl{W}})
    \text{Tr}\Big\{
    \Gamma_{\h{v}^n \h{\ul{u}}^n \h{w}_{V}^n 
    \h{\ulinew}_{U}^n}
    \!\!\!
    \sum_{\substack{(\ul{u}^n,  \ul{v}^n) \in T_{8\delta}^{(n)}\!\!}}\!\!\!
    \mcl{P}(\mcl{U})
    \\ & \hspace{18pt} \times\!
    \Big[
    \11_{\{\hat{m} \neq m_1\}}
    \Big(
    \frac{1}{1+4\tau^2}
    \tilde{\rho}_{\ul{v}^n 
    \ul{u}^n} +  N_{\tau}(\tilde{\rho}_{\ul{v}^n \ul{u}^n})
    \Big)
    + 
    \11_{\{\hat{m} = m_1\}}
    \Big(\frac{1+\tau^2}{1+4\tau^2}
    \mathcal{T}_{\hat{w}^n_{V}; \tau}^\mathrm{V_1}(\tilde{\rho}_{\ul{v}^n 
    \ul{u}^n}) + N_{\hat{w}^n_{V}; \tau}^\mathrm{V_1}(\tilde{\rho}_{\ul{v}^n \ul{u}^n})
    \Big)
    \Big]
    \Big\}, \\
    &\overset{(e)}{\leq} \sum_{\substack{\ul{m} \\ \ul{a}}} 
    \sum_{\substack{\h{v}^n\h{\ul{u}}^n\\ \h{w}_{V}^n 
    \h{\ulinew}_{U}^n}} 
    \sum_{\h{a}}
    \sum_{\substack{\h{m} \neq m^\oplus \\ \h{m}_3 \neq m_3 \\ \h{m}_4 \neq m_4}}
    \frac{p_{\ul{M}}(\ul{m})}{q^{3n}}
    \mcl{P}(\h{\mcl{U}})
    \mcl{P}(\h{\mcl{W}})
    2^{n(H(V_1,V_2)+\delta_1)}\sum_{v^n} p_{V}^n(v^n)
    \bigg[\text{Tr}\Big\{
    \Gamma_{\h{v}^n \h{\ul{u}}^n \h{w}_{V}^n 
    \h{\ulinew}_{U}^n} 
    \\& \hspace{20pt}\times
    \sum_{\substack{\ul{u}^n}}
    \mcl{P}(\mcl{U})
    \Big[\11_{\{\hat{m} \neq m_1\}}
    \sum_{\ul{v}^n : (\ulineu^n,\ulinev^n)\in T_{8\delta}^{(n)}}p^n_{\ulineV|V}(\ulinev^n|v^n) \bbm{1}_{\{v^n = v_1^n\oplus v_2^n \}}
    \tilde{\rho}_{\ul{v}^n \ul{u}^n} +
    \\
    & \hspace{40pt}+ 
    \11_{\{\hat{m} = m_1\}}
    \mathcal{T}_{\h{w}_{V}^n; \tau}^\mathrm{V_1}
   ~(\!\!\!\sum_{\ul{v}^n : (\ulineu^n,\ulinev^n)\in T_{8\delta}^{(n)}}\!\!\!p^n_{\ulineV|V}(\ulinev^n|v^n) \bbm{1}_{\{v^n = v_1^n\oplus v_2^n \}}
    \tilde{\rho}_{\ul{v}^n \ul{u}^n})
    \Big]
    \Big\} + \frac{16\sqrt{2}\tau |\mcl{H}_Z|^n}{\sqrt{\mcl{|W|}^n}} \bigg],\\
    &\overset{(f)}{=} \sum_{\substack{\ul{m} \\ \ul{a}}} 
     \sum_{\substack{\h{v}^n\h{\ul{u}}^n\\ \h{w}_{V}^n 
    \h{\ulinew}_{U}^n}} 
    \sum_{\h{a}}
    \sum_{\substack{\h{m} \neq m^\oplus  \\ \h{m}_3 \neq m_3 \\ \h{m}_4 \neq m_4}}
    \frac{p_{\ul{M}}(\ul{m})}{q^{3n}}
    \mcl{P}(\h{\mcl{U}})
    \mcl{P}(\hat{\mcl{W}})
    2^{n(H(V_1,V_2)+\delta_1)}
    \bigg[
    \text{Tr}\Big\{
    \Gamma_{\h{v}^n \h{\ul{u}}^n \h{w}_{V}^n 
    \h{\ulinew}_{U}^n}
    \\ & 
    \hspace{20pt}\times 
    \sum_{v^n \ulineu^n} p_{V}^n(v^n)
    \mcl{P}(\mcl{U})
    \Big[\11_{\{\hat{m}\neq m_1\}} \tilde{\rho}_{{v}^n \ul{u}^n} + \11_{\{\hat{m}= m_1\}}
    \mathcal{T}_{\h{w}_{V}^n; \tau}^\mathrm{V_1}
    (\tilde{\rho}_{{v}^n \ul{u}^n})
    \Big]
    \Big\} 
     + \frac{16\sqrt{2}\tau |\mcl{H}_Z|^n}{\sqrt{\mcl{|W|}^n}} \bigg], \\
    &\overset{(g)}{\leq} \sum_{\substack{\ul{m} \\ \ul{a}}} 
     \sum_{\substack{\h{v}^n\h{\ul{u}}^n\\ \h{w}_{V}^n 
    \h{\ulinew}_{U}^n}} 
    \sum_{\h{a}}
    \sum_{\substack{\h{m} \neq m^\oplus \\ \h{m}_3 \neq m_3 \\ \h{m}_4 \neq m_4}}
    \frac{p_{\ul{M}}(\ul{m})}{q^{3n}}
    \mcl{P}(\h{\mcl{U}})
    \mcl{P}(\hat{\mcl{W}})
    2^{n(H(V_1,V_2)+\delta_1)}
     \bigg[\text{Tr}\Big\{ {\Pi}^{\mathrm{V\ulineU}}_{\hat{v}^n\hat{\ulineu}^n}\rho
    \Big\}  + \frac{16\sqrt{2}\tau |\mcl{H}_Z|^n}{\sqrt{\mcl{|W|}^n}} \bigg],\\
    &\overset{(h)}{\leq}\sum_{\substack{\ul{m} \\ \ul{a}}}
    \sum_{\substack{\h{v}^n\h{\ul{u}}^n\\ \h{w}_{V}^n 
    \h{\ulinew}_{U}^n}} 
    \sum_{\h{a}}
    \sum_{\substack{\h{m} \neq m^\oplus \\ \h{m}_3 \neq m_3 \\ \h{m}_4 \neq m_4}}
    \frac{p_{\ul{M}}(\ul{m})}{q^{3n}}
    \mcl{P}(\h{\mcl{U}})
    \mcl{P}(\hat{\mcl{W}})
     2^{n(H(V_1,V_2)+\delta_1)} \bigg[2^{-n(I(VU_1U_2;Z) -\delta_1)} + \frac{16\sqrt{2}\tau |\mcl{H}_Z|^n}{\sqrt{\mcl{|W|}^n}}  \bigg],
    \\
    & \overset{(i)}{\leq} 
   2^{\Big\{n\Big[
    \Big(\frac{3k+l + l_1+l_2}{n}\Big)\log q-
    3\log q +
    H(V)+
    H(V_1,V_2)+3\delta_1
    -I(V, U_1, U_2;Z)_{\sigma}\Big]\Big\}},
\end{align*}
 
where $(a)$ follows by bounding $\11_{\mcl{A}} \leq 1$, $(b)$ follows by using $\mcl{P}(\mcl{V},\hat{\mcl{V}}) = \frac{1}{q^{3n}}$ and rearranging the terms, $(c)$ follows by using the fact that $w_V$ used by the decoder is identical to  $w_{V_1}$ and expanding $\CalP(\CalW|\hat{\CalW})$ (for $\hat{m} \neq m^{\oplus},\hat{m}_3 \neq m_3,\hat{m}_4 \neq m_4$) as follows:
\begin{align*}
    \sum_{\ulinew^n}\CalP(\CalW|\hat{\CalW}) &= \sum_{\ulinew^n} \CalP(\ulineW^n(\ulinem) = \ulinew^n|{W}_{V}^n(\hat{m}) = \hat{w}_V^n,{\ulineW}_{U}^n(\hat{m}_3,\hat{m}_4) = \hat{w}_{U}^n),\\
   & = \begin{cases} 
      \sum_{\ulinew^n} 1/|\CalW|^{4n} & \colon \hat{m}\neq m_1 \\
      \sum_{\ulinew^n} \11_{\{w_{V_1}^n = \hat{w}_{V}^n\}} 1/|\CalW|^{3n}  & \colon \hat{m} = m_1,
   \end{cases}
\end{align*}
(d) follows from the observations \cite[Section 4]{sen2021unions}:
\begin{align*}
    \sum_{\ulinew^n} \frac{1}{|\CalW|^{4n}}\CalT^{\mathrm{V\ulineU}}_{\ulinew^n;\tau}(\tilde{\rho}_{\ulinev^n\ulineu^n}) &= \frac{1}{1+4\tau^2}\tilde{\rho}_{\ulinev^n\ulineu^n} + N_{\tau}(\tilde{\rho}_{\ulinev^n\ulineu^n}),\\
    \sum_{\ulinew^n} \11_{\{w_{V_1}^n=\hat{w}_V^n\}} \frac{1}{|\CalW|^{3n}}\CalT^{\mathrm{V\ulineU}}_{\ulinew^n;\tau}(\tilde{\rho}_{\ulinev^n\ulineu^n})
    &= \frac{1+\tau^2}{1+4\tau^2}\CalT^{\mathrm{V_1}}_{\hat{w}_V^n;\tau}(\tilde{\rho}_{\ulinev^n\ulineu^n}) + N^{\mathrm{V_1}}_{\hat{w}_V^n;\tau}(\tilde{\rho}_{\ulinev^n\ulineu^n}),
\end{align*}
(e) follows from the typicality property that for $\ulinev^n\in T_{8\delta}^{(n)}(p_{V_1V_2})$ and sufficiently large $n$, we have $\sum_{v^n}p_V^n(v^n) p_{\ulineV|V}^n(\ulinev^n|v^n) \leq 2^{-n(H(V_1,V_2)+\delta_1)}$, and the following observations found in \cite[Section 4]{sen2021unions}: 
\begin{align*}
   (i) \;\norm{N_{\tau}(\tilde{\rho}_{v^n\ulineu^n})}_\infty & \leq 4\sqrt{2}\tau/\sqrt{|\CalW|^n},\quad 
   (ii) \;||N^{\mathrm{V_1}}_{\hat{w}_V^n;\tau}(\tilde{\rho}_{v^n\ulineu^n})||_\infty \leq4\sqrt{2}\tau/\sqrt{|\CalW|^n},\\ & \;\;\; (iii)\; 
    ||\Gamma_{\h{v}^n \h{\ul{u}}^n \h{w}_{V}^n 
\h{\ulinew}_{U}^n}||_1 \leq 2|\CalH_Z|^n,
\end{align*}
(f) follows by using the definition $\tilde{\rho}_{v^n\ulineu^n} = \sum_{\ulinev^n} p_{\ulineV|V}^n(\ulinev^n|v^n) \11_{\{v^n = v_1^n \oplus v_2^n\}} {\rho}_{\ulinev^n\ulineu^n} \tensor \ovec$,
(g) follows from \cite[Equation 8]{sen2021unions}, 
and the fact that $\tr\{\bar{\Pi}^{\mathrm{V\ulineU}}_{\hat{v}^n\hat{\ulineu}^n}\tilde{\rho}\} = 
\tr\{{\Pi}^{\mathrm{V\ulineU}}_{\hat{v}^n\hat{\ulineu}^n}\rho\}$, 
(h) follows from Proposition \ref{prop:povmproperty}, and finally (i) follows by choosing $|\mcl{W}| \leq 2^{I(V,U_1,U_2;Z)_\sigma}$.
\subsubsection{Analysis of $T_{2\ulineU}$}

We now analyze the error event $T_{2\ulineU}$ using similar techniques as used for analyzing $T_{2V\ulineU}$. 
Define the following events for 
${m}^{\oplus}$, $\h{m}_3, \h{m}_4$ and ${a}^{\oplus}$:
\[ \h{\mcl{V}} \deq \{{V}^n({a}^{\oplus},{m}^{\oplus}) = {v}^n\},  
\quad \h{\mcl{U}} \triangleq {\{
U_j^n(\h{m}_{j+2}) = \h{u}_j^n \ : \ j \in [2]
\}},\]
\[\hat{\mcl{W}} \deq \{
{W}_{V}^n({m}^{\oplus}) = {w}_{V}^n,~{\ulineW}_{U}^n(\h{m}_3,\h{m}_4) = \hat{\ulinew}_{U}^n
\}.\]
\begin{align*}
    &\mathbb{E}_{\mathcal{P}}\left[T_{2\ulineU}\right]= \mathbb{E} \Bigg[ \sum_{\substack{\ul{m} \\ \ul{a}}} 
    \sum_{\substack{ v^n, \h{\ul{u}}^n, \ul{v}^n, \ul{u}^n\\ w_V^n,\ulinew^n, 
    \h{\ulinew}_{U}^n}} 
    \sum_{\substack{ \h{m}_3 \neq m_3 \\ \h{m}_4 \neq m_4}}
    p_{\ul{M}}(\ul{m})
    \text{Tr}\left\{
    \Gamma_{{v}^n \h{\ul{u}}^n {w}_{V}^n 
    \h{\ulinew}_{U}^n}
    \rho'_{\ul{v}^n 
    \ul{u}^n
    \ulinew^n}
    \right\}
    \bbm{1}_{\mcl{V}}
    \bbm{1}_{\h{\mcl{V}}}
    \bbm{1}_{\mcl{U}}
    \bbm{1}_{\h{\mcl{U}}}
    \bbm{1}_{\mcl{W}}
    \bbm{1}_{\h{\mcl{W}}}
    \bbm{1}_{\mcl{A}} 
    \bbm{1}_{\mcl{E}} 
    \Bigg],  \\
    &\overset{(a)}{\leq} 
    \sum_{\substack{\ul{m} \\ \ul{a}}}
    \sum_{\substack{v^n(\ul{u}^n, \ul{v}^n) \in T_{8\delta}^{(n)}\\ \h{\ul{u}}^n \ulinew^n  w_V^n
    \h{\ulinew}_{U}^n}} 
    \sum_{\substack{\h{m}_3 \neq m_3 \\ \h{m}_4 \neq m_4}}
    p_{\ul{M}}(\ul{m})
    \text{Tr}\left\{
    \Gamma_{{v}^n \h{\ul{u}}^n {w}_{V}^n 
    \h{\ulinew}_{U}^n}
    \rho'_{\ul{v}^n 
    \ul{u}^n
    \ulinew^n}
    \right\}
    \mcl{P}(\mcl{V},\h{\mcl{V}})
    \mcl{P}(\mcl{U})
    \mcl{P}(\h{\mcl{U}})
    \mcl{P}(\mcl{W},\h{\mcl{W}}), \\
    &\overset{(b)}{=} 
    \sum_{\substack{\ul{m} \\ \ul{a}}}
    \sum_{\substack{v^n\ul{v}^n\h{\ul{u}}^n\\ w_V^n
    \h{\ulinew}_{U}^n}} 
    \sum_{\substack{\\ \h{m}_3 \neq m_3 \\ \h{m}_4 \neq m_4}}
    \frac{p_{\ul{M}}(\ul{m})}{q^{2n}}
    \mcl{P}(\h{\mcl{U}})
    \mcl{P}(\h{\mcl{W}})
    \text{Tr}\Big\{
    \Gamma_{{v}^n \h{\ul{u}}^n {w}_{V}^n 
    \h{\ulinew}_{U}^n}
    \sum_{\substack{\ul{u}^n: (\ul{u}^n,  \ul{v}^n) \in T_{8\delta}^{(n)}\!\! \\ \ul{w}^n}} 
    \mcl{P}(\mcl{U})
    \mcl{P}(\mcl{W}|\h{\mcl{W}})
    \rho'_{\ul{v}^n 
    \ul{u}^n
    \ul{w}^n}
    \Big\}, \\
    &\overset{(c)}{=} \sum_{\substack{\ul{m} \\ \ul{a}}} 
    \sum_{\substack{v^n\ul{v}^n\h{\ul{u}}^n\\w_V^n
    \h{\ulinew}_{U}^n}} 
    \sum_{\substack{  \\ \h{m}_3 \neq m_3 \\ \h{m}_4 \neq m_4}}
    \frac{p_{\ul{M}}(\ul{m})}{q^{2n}}
    \mcl{P}(\h{\mcl{U}})
    \mcl{P}(\h{\mcl{W}})
    \text{Tr}\Big\{
    \Gamma_{{v}^n \h{\ul{u}}^n {w}_{V}^n 
    \h{\ulinew}_{U}^n} \!\!\!
    \sum_{\substack{\ul{u}^n:(\ul{u}^n,  \ul{v}^n) \in T_{8\delta}^{(n)}\!\!}} \!\!\!
    \mcl{P}(\mcl{U})
    \\ & \hspace{30pt} \times 
    \Big[ 
    \11_{\{m^{\oplus} \neq m_1\}}
    \sum_{ \ulinew^n} \frac{1}{|\mcl{W}|^{4n}}
    \mathcal{T}_{\ulinew^n; \tau}^\mathrm{\ulineV\ulineU}(\tilde{\rho}_{\ul{v}^n 
    \ul{u}^n})
    +
     \11_{\{m^{\oplus} = m_1\}}
    \sum_{\ulinew^n}
    \11_{\{w_{V_1}^n = w_{V}^n\}}
    \frac{1}{|\mcl{W}|^{3n}}
    \mathcal{T}_{\ulinew^n; \tau}^\mathrm{\ulineV\ulineU}(\tilde{\rho}_{\ul{v}^n 
    \ul{u}^n})
    \Big]
    \Big\},
    \\
    &\overset{(d)}{\leq} \sum_{\substack{\ul{m} \\ \ul{a}}} 
    \sum_{\substack{v^n \ul{v}^n\h{\ul{u}}^n\\ w_V^n
    \h{\ulinew}_{U}^n}} 
    \sum_{\substack{ \h{m}_3 \neq m_3 \\ \h{m}_4 \neq m_4}}
    \frac{p_{\ul{M}}(\ul{m})}{q^{2n}}
    \mcl{P}(\h{\mcl{U}})
    \mcl{P}(\h{\mcl{W}})
    \text{Tr}\Big\{
    \Gamma_{{v}^n \h{\ul{u}}^n {w}_{V}^n 
    \h{\ulinew}_{U}^n} 
    \sum_{\substack{\ul{u}^n: (\ul{u}^n,  \ul{v}^n) \in T_{8\delta}^{(n)}\!\!}}
    \mcl{P}(\mcl{U})
    \\ & \hspace{30pt} \times 
    \Big[
    \11_{\{m^{\oplus}\neq m_1\}}
    \Big(
    \tilde{\rho}_{\ul{v}^n 
    \ul{u}^n} + N_{\tau}(\tilde{\rho}_{\ul{v}^n 
    \ul{u}^n})
    \Big)+
    \11_{\{m^{\oplus}  = m_1\}}
    \Big(
    \frac{1+\tau^2}{1+4\tau^2}
    \mathcal{T}_{w^n_{V}; \tau}^\mathrm{V_1}(\tilde{\rho}_{\ul{v}^n 
    \ul{u}^n}) + N_{w^n_{V}; \tau}^\mathrm{V_1}(\tilde{\rho}_{\ul{v}^n 
    \ul{u}^n}) \Big)
    \Big]
    \Big\}, \\
    &\overset{(e)}{\leq} \sum_{\substack{\ul{m} \\ \ul{a}}} 
     \sum_{\substack{\h{\ul{u}}^n\\ {w}_{V}^n 
    \h{\ulinew}_{U}^n}} 
    \sum_{\substack{\h{m}_3 \neq m_3 \\ \h{m}_4 \neq m_4}}
    \frac{p_{\ul{M}}(\ul{m})}{q^{2n}}
    \mcl{P}(\h{\mcl{U}})
    \mcl{P}(\hat{\mcl{W}})
    2^{n(H(V_1,V_2)+\delta_1)}
    \sum_{v^n} p_{V}^n(v^n)
    \Bigg[
    \text{Tr}\Big\{
    \Gamma_{{v}^n \h{\ul{u}}^n {w}_{V}^n 
    \h{\ulinew}_{U}^n}  
    \sum_{\substack{\ul{u}^n }}
    \mcl{P}(\mcl{U})
    \\ & \hspace{30pt}\times 
    \Big[
    \11_{\{m^{\oplus} \neq m_1 \}}
    p^n_{\ulineV|V}(\ulinev^n|v^n) \bbm{1}_{\{v^n = v_1^n\oplus v_2^n \}}
    \tilde{\rho}_{\ul{v}^n 
    \ul{u}^n}
    \\ & \hspace{50pt}+
    \11_{\{m^{\oplus} = m_1\}}
    \mathcal{T}_{{w}_{V}^n; \tau}^\mathrm{V_1}
   \ (\!\!\!\sum_{\ul{v}^n : (\ulineu^n,\ulinev^n)\in T_{8\delta}^{(n)}} \!\!\!p^n_{\ulineV|V}(\ulinev^n|v^n) \bbm{1}_{\{v^n = v_1^n\oplus v_2^n \}}
    \tilde{\rho}_{\ul{v}^n 
    \ul{u}^n}
    )
    \Big]
    \Big\} 
    + \frac{16\sqrt{2}\tau |\mcl{H}_Z|^n}{\sqrt{\mcl{|W|}^n}}     \Bigg], \\
    &\overset{(f)}{=} \sum_{\substack{\ul{m} \\ \ul{a}}} 
     \sum_{\substack{\h{\ul{u}}^n\\ {w}_{V}^n 
    \h{\ulinew}_{U}^n}} 
    \sum_{\substack{\h{m}_3 \neq m_3 \\ \h{m}_4 \neq m_4}}
    \frac{p_{\ul{M}}(\ul{m})}{q^{2n}}
    \mcl{P}(\h{\mcl{U}})
    \mcl{P}(\hat{\mcl{W}})
    2^{n(H(V_1,V_2)+\delta_1)}
    \sum_{v^n} p_{V}^n(v^n)
    \Bigg[ 
    \text{Tr}\Big\{
    \Gamma_{{v}^n \h{\ul{u}}^n {w}_{V}^n 
    \h{\ulinew}_{U}^n} 
    \\ & \hspace{30pt}\times
    \Big[
    \11_{\{m^\oplus \neq m_1\}}
   \sum_{\substack{\ul{u}^n }}
    \mcl{P}(\mcl{U})
    \tilde{\rho}_{{v}^n 
    \ulineu^n}+
    \11_{\{m^\oplus = m_1\}}
    \mathcal{T}_{\h{w}_{V}^n; \tau}^\mathrm{V_1}
    (\sum_{\substack{\ul{u}^n }}
    \mcl{P}(\mcl{U}) \tilde{\rho}_{{v}^n 
    \ulineu^n}) 
    \Big]
    \Big\} + \frac{16\sqrt{2}\tau |\mcl{H}_Z|^n}{\sqrt{\mcl{|W|}^n}} \Bigg], \\
    &\overset{(g)}{\leq} \sum_{\substack{\ul{m} \\ \ul{a}}} 
     \sum_{\substack{{w}_{V}^n 
    \h{\ulinew}_{U}^n}} 
    \sum_{\substack{\h{m}_3 \neq m_3 \\ \h{m}_4 \neq m_4}} \!\!
    \frac{p_{\ul{M}}(\ul{m})}{q^{2n}}
    \mcl{P}(\hat{\mcl{W}})
    2^{n(H(V_1,V_2)+\delta_1)}\sum_{v^n}
    p_{V}^n(v^n) 
    \Bigg[\sum_{\h{\ul{u}}^n}\mcl{P}(\h{\mcl{U}})
    \text{Tr}\Big\{ \! {\Pi}^{\mathrm{V}}_{{v}^n\hat{\ulineu}^n}
    {\rho}_{v^n}
    \! \Big \} \! +\! \frac{16\sqrt{2}\tau |\mcl{H}_Z|^n}{\sqrt{\mcl{|W|}^n}} \Bigg],\\
    &\overset{(h)}{\leq}\sum_{\substack{\ul{m} \\ \ul{a}}} 
    \sum_{\substack{{w}_{V}^n 
    \h{\ulinew}_{U}^n}} 
    \sum_{\substack{\h{m}_3 \neq m_3 \\ \h{m}_4 \neq m_4}}
    \frac{p_{\ul{M}}(\ul{m})}{q^{2n}}
    \mcl{P}(\hat{\mcl{W}})
    2^{n(H(V_1,V_2)+\delta_1)} \Big[2^{-n(I(U_1U_2;Z|V) -\delta_1)} + \frac{16\sqrt{2}\tau |\mcl{H}_Z|^n}{\sqrt{\mcl{|W|}^n}}\Big],
    \\
    & \overset{(i)}{\leq} 
   2^{\left\{n\left[
    \left(\frac{2k+ l_1+l_2}{n}\right)\log q-
    2\log q +
    H(V_1,V_2)+3\delta_1
    -I( U_1, U_2;Z|V)_{\sigma}\right]\right\}}. 
\end{align*}
  The above sequence of steps is analogous to the above steps used for deriving an upper bound on $T_{2V\ulineU}$ and follow from the same set of arguments.
  
Similarly, for $i,j \in [2]$ and $i\neq j$, we get:
\begin{align*}
    \EE_{\CalP}[T_{2VU_j}] &\leq 2^{\left\{n\left[
    \left(\frac{3k + l + l_j}{n}\right)\log q-
    3\log q +
    H(V_1,V_2) + H(V) + 3\delta_1
    -I(V, U_j;Z|U_i)_{\sigma}\right]\right\}},\\
    \EE_{\CalP}[T_{2V}] &\leq 2^{\left\{n\left[
    \left(\frac{3k+l}{n}\right)\log q-
    3\log q +
    H(V_1,V_2) + H(V) + 3\delta_1
    -I(V;Z|U_1,U_2)_{\sigma}\right]\right\}},\\
    \EE_{\CalP}[T_{2U_j}] &\leq 2^{\left\{n\left[
    \left(\frac{2k + l_j}{n}\right)\log q-
    2\log q +
    H(V_1,V_2)+3\delta_1
    -I(U_j;Z|V,U_i)_{\sigma} \right]\right\}}.
\end{align*}
This completes the proof of the Proposition \ref{prop:erroranalysis}.

\newpage
\bibliographystyle{IEEEtran}
\bibliography{IEEEabrv,references}

\end{document}